%
%
%
\expandafter \def \csname CHAPLABELintro\endcsname {1}
\expandafter \def \csname EQLABELintegrals\endcsname {1.1?}
\expandafter \def \csname EQLABELquintic\endcsname {1.2?}
\expandafter \def \csname EQLABELyukone\endcsname {1.3?}
\expandafter \def \csname EQLABELyuktwo\endcsname {1.4?}
\expandafter \def \csname EQLABELmmap\endcsname {1.5?}
\expandafter \def \csname EQLABELPF\endcsname {1.6?}
\expandafter \def \csname EQLABELperiods\endcsname {1.7?}
\expandafter \def \csname EQLABELsemiperiod\endcsname {1.8?}
\expandafter \def \csname EQLABELfirstresult\endcsname {1.9?}
\expandafter \def \csname CHAPLABELprelims\endcsname {2}
\expandafter \def \csname EQLABELfactone\endcsname {2.1?}
\expandafter \def \csname EQLABELfacttwo\endcsname {2.2?}
\expandafter \def \csname EQLABELNseries\endcsname {2.3?}
\expandafter \def \csname FIGLABELdisk\endcsname {2.1?}
\expandafter \def \csname EQLABELp-order\endcsname {2.4?}
\expandafter \def \csname EQLABELsigdef\endcsname {2.5?}
\expandafter \def \csname EQLABELap\endcsname {2.6?}
\expandafter \def \csname EQLABELgammadef\endcsname {2.7?}
\expandafter \def \csname EQLABELnfac\endcsname {2.8?}
\expandafter \def \csname EQLABELgammaformula\endcsname {2.9?}
\expandafter \def \csname EQLABELclassicalref\endcsname {2.10?}
\expandafter \def \csname EQLABELpremult\endcsname {2.11?}
\expandafter \def \csname EQLABELpremultclass\endcsname {2.12?}
\expandafter \def \csname EQLABELmultclass\endcsname {2.13?}
\expandafter \def \csname FIGLABELsquare\endcsname {2.2?}
\expandafter \def \csname FIGLABELcube\endcsname {2.3?}
\expandafter \def \csname EQLABELacongruence\endcsname {2.14?}
\expandafter \def \csname EQLABELFrobzero\endcsname {2.15?}
\expandafter \def \csname CHAPLABELallperiods\endcsname {3}
\expandafter \def \csname EQLABELautomorphisms\endcsname {3.1?}
\expandafter \def \csname EQLABELgenericperiod\endcsname {3.2?}
\expandafter \def \csname FIGLABELloci\endcsname {3.1?}
\expandafter \def \csname EQLABELexact\endcsname {3.3?}
\expandafter \def \csname EQLABELcaseoneid\endcsname {3.4?}
\expandafter \def \csname EQLABELPF\endcsname {3.5?}
\expandafter \def \csname EQLABELFfrob\endcsname {3.6?}
\expandafter \def \csname EQLABELAm\endcsname {3.7?}
\expandafter \def \csname FIGLABELloopandwedge\endcsname {3.2?}
\expandafter \def \csname EQLABELupsilonzero\endcsname {3.8?}
\expandafter \def \csname EQLABELLv\endcsname {3.9?}
\expandafter \def \csname EQLABELgencoeffs\endcsname {3.10?}
\expandafter \def \csname CHAPLABELcalczero\endcsname {4}
\expandafter \def \csname EQLABELfacttwo\endcsname {4.1?}
\expandafter \def \csname EQLABELnmcondition\endcsname {4.2?}
\expandafter \def \csname EQLABELpiofalpha\endcsname {4.3?}
\expandafter \def \csname EQLABELsimplecase\endcsname {4.4?}
\expandafter \def \csname CHAPLABELcalcone\endcsname {5}
\expandafter \def \csname EQLABELnewnu\endcsname {5.1?}
\expandafter \def \csname TABLABELvtableone\endcsname {5.1?}
\expandafter \def \csname EQLABELpoch\endcsname {5.2?}
\expandafter \def \csname EQLABELvcontribs\endcsname {5.3?}
\expandafter \def \csname EQLABELdiscrepant\endcsname {5.4?}
\expandafter \def \csname EQLABELorderone\endcsname {5.5?}
\expandafter \def \csname CHAPLABELhigher\endcsname {6}
\expandafter \def \csname EQLABELAm\endcsname {6.1?}
\expandafter \def \csname EQLABELAmfour\endcsname {6.2?}
\expandafter \def \csname EQLABELfk\endcsname {6.3?}
\expandafter \def \csname EQLABELNpsi\endcsname {6.4?}
\expandafter \def \csname EQLABELhratio\endcsname {6.5?}
\expandafter \def \csname EQLABELaovera\endcsname {6.6?}
\expandafter \def \csname EQLABELpadicN\endcsname {6.7?}
\expandafter \def \csname EQLABELbetam\endcsname {6.8?}
\expandafter \def \csname EQLABELAminfty\endcsname {6.9?}
\expandafter \def \csname EQLABELlimit\endcsname {6.10?}
\expandafter \def \csname EQLABELBoyarski\endcsname {6.11?}
\expandafter \def \csname EQLABELDworkfn\endcsname {6.12?}
\expandafter \def \csname EQLABELDworkcoeffs\endcsname {6.13?}
\expandafter \def \csname EQLABELAcoeffs\endcsname {6.14?}
\expandafter \def \csname EQLABELnumericalbeta\endcsname {6.15?}
\expandafter \def \csname EQLABELends\endcsname {6.16?}
\expandafter \def \csname EQLABELshortbeta\endcsname {6.17?}
\expandafter \def \csname EQLABELforcalc\endcsname {6.18?}
\expandafter \def \csname EQLABELfinalN\endcsname {6.19?}
\expandafter \def \csname EQLABELnus\endcsname {6.20?}
\expandafter \def \csname EQLABELbetas\endcsname {6.21?}
\expandafter \def \csname CHAPLABELzerodim\endcsname {7}
\expandafter \def \csname EQLABELreclaw\endcsname {7.1?}
\expandafter \def \csname EQLABELperiod\endcsname {7.2?}
\expandafter \def \csname EQLABELprenpsi\endcsname {7.3?}
\expandafter \def \csname EQLABELnpsi\endcsname {7.4?}
\expandafter \def \csname EQLABELmpsi\endcsname {7.5?}
\expandafter \def \csname CHAPLABELcubics\endcsname {8}
\expandafter \def \csname EQLABELAcubic\endcsname {8.1?}
\expandafter \def \csname EQLABEL\endcsname {8.2?}
\expandafter \def \csname CHAPLABELgauss\endcsname {9}
\expandafter \def \csname EQLABELseries\endcsname {9.1?}
\expandafter \def \csname EQLABELgausssums\endcsname {9.2?}
\expandafter \def \csname EQLABELinversion\endcsname {9.3?}
\expandafter \def \csname EQLABELGKformula\endcsname {9.4?}
\expandafter \def \csname EQLABELyPeqn\endcsname {9.5?}
\expandafter \def \csname EQLABELcharcases\endcsname {9.6?}
\expandafter \def \csname EQLABELintermediategauss\endcsname {9.7?}
\expandafter \def \csname EQLABELnuexpr\endcsname {9.8?}
\expandafter \def \csname EQLABELyg\endcsname {9.9?}
\expandafter \def \csname EQLABELssum\endcsname {9.10?}
\expandafter \def \csname TABLABELvtabletwo\endcsname {9.1?}
\expandafter \def \csname EQLABELlsum\endcsname {9.11?}
\expandafter \def \csname EQLABELnuY\endcsname {9.12?}
\expandafter \def \csname EQLABELnewnewnu\endcsname {9.13?}
\expandafter \def \csname EQLABELcoeffrels\endcsname {9.14?}
\expandafter \def \csname EQLABELprecnf\endcsname {9.15?}
\expandafter \def \csname EQLABELgammamult\endcsname {9.16?}
\expandafter \def \csname EQLABELnewcoefrels\endcsname {9.17?}
\expandafter \def \csname EQLABELgaussqsums\endcsname {9.18?}
\expandafter \def \csname EQLABELnewnewnuq\endcsname {9.19?}
\expandafter \def \csname CHAPLABELintro\endcsname {10}
\expandafter \def \csname EQLABELpolydef\endcsname {10.1?}
\expandafter \def \csname EQLABELZdef\endcsname {10.2?}
\expandafter \def \csname EQLABELzetaM\endcsname {10.3?}
\expandafter \def \csname EQLABELNdecomp\endcsname {10.4?}
\expandafter \def \csname EQLABELZrat\endcsname {10.5?}
\expandafter \def \csname EQLABELQrel\endcsname {10.6?}
\expandafter \def \csname EQLABELzetaW\endcsname {10.7?}
\expandafter \def \csname EQLABELlcsl\endcsname {10.8?}
\expandafter \def \csname CHAPLABELEuler\endcsname {11}
\expandafter \def \csname EQLABELRAB\endcsname {11.1?}
\expandafter \def \csname EQLABELEcurve\endcsname {11.2?}
\expandafter \def \csname TABLABELmonomials\endcsname {11.1?}
\expandafter \def \csname EQLABELisomorphisms\endcsname {11.3?}
\expandafter \def \csname EQLABELsymone\endcsname {11.4?}
\expandafter \def \csname EQLABELhyperelliptic\endcsname {11.5?}
\expandafter \def \csname EQLABELmap\endcsname {11.6?}
\expandafter \def \csname EQLABELsymtwo\endcsname {11.7?}
\expandafter \def \csname EQLABELEFrels\endcsname {11.8?}
\expandafter \def \csname EQLABELsingids\endcsname {11.9?}
\expandafter \def \csname EQLABELZEuler\endcsname {11.10?}
\expandafter \def \csname EQLABELR\endcsname {11.11?}
\expandafter \def \csname EQLABELformzeta\endcsname {11.12?}
\expandafter \def \csname CHAPLABELZfn\endcsname {12}
\expandafter \def \csname EQLABELpaireqs\endcsname {12.1?}
\expandafter \def \csname TABLABELpsiequalszero\endcsname {12.1?}
\expandafter \def \csname FIGLABELSthreefig\endcsname {12.1?}
\expandafter \def \csname FIGLABELStwofig\endcsname {12.2?}
\expandafter \def \csname CHAPLABELgeneral\endcsname {13}
\expandafter \def \csname CHAPLABELmirrorquintic\endcsname {14}
\expandafter \def \csname FIGLABELtwoface\endcsname {14.1?}
\expandafter \def \csname EQLABELmirrorquintic\endcsname {14.1?}
\expandafter \def \csname EQLABELmirrorquintic\endcsname {14.2?}
\expandafter \def \csname TABLABELcontributions\endcsname {14.1?}
\expandafter \def \csname EQLABELn0\endcsname {14.3?}
\expandafter \def \csname EQLABELnzero\endcsname {14.4?}
\expandafter \def \csname EQLABELn0eq\endcsname {14.5?}
\expandafter \def \csname EQLABELprenstar\endcsname {14.6?}
\expandafter \def \csname EQLABELnstar\endcsname {14.7?}
\expandafter \def \csname FIGLABELflop\endcsname {14.2?}
\expandafter \def \csname CHAPLABELmodfive\endcsname {15}
\expandafter \def \csname CHAPLABELquestions\endcsname {16}
\expandafter \def \csname CHAPLABELA\endcsname {17}
\expandafter \def \csname EQLABELsingP\endcsname {17.1?}
\expandafter \def \csname CHAPLABELB\endcsname {18}
\expandafter \def \csname EQLABELFper\endcsname {18.1?}
\expandafter \def \csname EQLABELmirrormatrix\endcsname {18.2?}
\expandafter \def \csname EQLABELocticP\endcsname {18.3?}
\expandafter \def \csname EQLABELDrels\endcsname {18.4?}
\expandafter \def \csname EQLABELmirrormatrix\endcsname {18.5?}
%
%
\magnification=1200

\font\eightrm=cmr8 at 8pt
\font\fourteenrm=cmr12 at 14pt
\font\seventeenrm=cmr17 at 17pt
\font\twentyonerm=cmr17 at 21pt

\font\ss=cmss10

\font\csc=cmcsc10

\font\twelvecal=cmsy10 at 12pt

\font\twelvemath=cmmi12

\font\fourteenbold=cmbx12 at 14pt
\font\seventeenbold=cmbx7 at 17pt

\font\fively=lasy5
\font\sevenly=lasy7
\font\tenly=lasy10

\textfont10=\tenly
\scriptfont10=\sevenly    
\scriptscriptfont10=\fively
\parskip=10pt
\parindent=20pt
\def\today{\ifcase\month\or January\or February\or March\or April\or May\or June
       \or July\or August\or September\or October\or November\or December\fi
       \space\number\day, \number\year}

\def\title#1{\footline={\ifnum\pageno<2\hfil
       \else\hss\tenrm\folio\hss\fi}\vskip1truein\centerline{{#1}   
       \footnote{\raise1ex\hbox{*}}{\eightrm Supported in part
       by the Robert A. Welch Foundation and N.S.F. Grants 
       PHY-880637 and\break PHY-8605978.}}}

\def\newpage{\vfill\eject}
\def\abstract#1{\centerline{\bf ABSTRACT}\vskip.2truein{\narrower\noindent#1
       \smallskip}}

\def\runninghead#1#2{\voffset=2\baselineskip\nopagenumbers
       \headline={\ifodd\pageno\rightheadline\else \leftheadline\fi}
       \def\rightheadline{{\sl#1}\hfill{\rm\folio}}
       \def\leftheadline{{\rm\folio}\hfill{\sl#2}}}
\def\SS{\mathhexbox278}

\newcount\footnoteno
\def\Footnote#1{\advance\footnoteno by 1
                \let\SF=\empty 
                \ifhmode\edef\SF{\spacefactor=\the\spacefactor}\/\fi
                $^{\the\footnoteno}$\ignorespaces
                \SF\vfootnote{$^{\the\footnoteno}$}{#1}}

\def\figbox#1#2#3{\vbox{\vskip15pt
                   \vbox{\hrule
                    \hbox{\vrule
                     \vbox{\vskip12truept\centerline #1 \vskip6truept
                          {\hskip.4truein\vbox{\hsize=5truein\noindent
                          {\bf Figure\hskip5truept#2:}\hskip5truept#3}}
                     \vskip18truept}
                    \vrule}
                   \hrule}}}
\def\place#1#2#3{\vbox to0pt{\kern-\parskip\kern-7pt
                             \kern-#2truein\hbox{\kern#1truein #3}
                             \vss}\nointerlineskip}
\def\figurecaption#1#2{\kern.75truein\vbox{\hsize=5truein\noindent{\bf Figure
    \figlabel{#1}:} #2}}
\def\tablecaption#1#2{\kern.75truein\lower12truept\hbox{\vbox{\hsize=5truein
    \noindent{\bf Table\hskip5truept\tablabel{#1}:} #2}}}
\def\boxed#1{\lower3pt\hbox{
                       \vbox{\hrule\hbox{\vrule
                         \vbox{\kern2pt\hbox{\kern3pt#1\kern3pt}\kern3pt}\vrule}
                         \hrule}}}
\def\a{\alpha}
\def\b{\beta}
\def\g{\gamma}\def\G{\Gamma}
\def\d{\delta}\def\D{\Delta}
\def\e{\epsilon}\def\ve{\varepsilon}
\def\z{\zeta}

\def\Th{\Theta}

\def\k{\kappa}
\def\l{\lambda}\def\L{\Lambda}
\def\m{\mu}
\def\n{\nu}
\def\x{\xi}

\def\p{\pi}\def\vp{\varpi}
\def\r{\rho}
\def\s{\sigma}

\def\ph{\phi}

\def\ps{\psi}

\def\ca#1{\relax\ifmmode {{\cal #1}}\else $\cal #1$\fi}

\def\calb{{\cal B}}

\def\calm{{\cal M}}

\def\inbar{\vrule height1.5ex width.4pt depth0pt}
\def\IB{\relax{\rm I\kern-.18em B}}
\def\IC{\relax\hbox{\kern.25em$\inbar\kern-.3em{\rm C}$}}
\def\ID{\relax{\rm I\kern-.18em D}}
\def\IE{\relax{\rm I\kern-.18em E}}
\def\IF{\relax{\rm I\kern-.18em F}}
\def\IG{\relax\hbox{\kern.25em$\inbar\kern-.3em{\rm G}$}}
\def\IH{\relax{\rm I\kern-.18em H}}
\def\II{\relax{\rm I\kern-.18em I}}
\def\IK{\relax{\rm I\kern-.18em K}}
\def\IL{\relax{\rm I\kern-.18em L}}
\def\IM{\relax{\rm I\kern-.18em M}}
\def\IN{\relax{\rm I\kern-.18em N}}
\def\IO{\relax\hbox{\kern.25em$\inbar\kern-.3em{\rm O}$}}
\def\IP{\relax{\rm I\kern-.18em P}}
\def\IQ{\relax\hbox{\kern.25em$\inbar\kern-.3em{\rm Q}$}}
\def\IR{\relax{\rm I\kern-.18em R}}
\def\IZ{\relax\ifmmode\hbox{\ss Z\kern-.4em Z}\else{\ss Z\kern-.4em Z}\fi}
\def\IGa{\relax{\rm I}\kern-.18em\Gamma}
\def\IPi{\relax{\rm I}\kern-.18em\Pi}
\def\ITh{\relax\hbox{\kern.25em$\inbar\kern-.3em\Theta$}}
\def\IOm{\relax\thinspace\inbar\kern1.95pt\inbar\kern-5.525pt\Omega}


\def\noblackboxes{\overfullrule=0pt}
\def\define{\buildrel\rm def\over =}

\def\cy{Calabi--Yau} 
\def\cym{Calabi--Yau manifold}
\def\cys{Calabi--Yau manifolds}

\def\K{K\"ahler}

\def\H#1#2{\relax\ifmmode {H^{#1#2}}\else $H^{#1 #2}$\fi}
\def\M{\relax\ifmmode{\calm}\else $\calm$\fi}

\def\Bigcheck{\lower3.8pt\hbox{\smash{\hbox{{\twentyonerm \v{}}}}}}
\def\bigboldcheck{\smash{\hbox{{\seventeenbold\v{}}}}}

\def\Bighat{\lower3.8pt\hbox{\smash{\hbox{{\twentyonerm \^{}}}}}}

\def\Msharp{\relax\ifmmode{\calm^\sharp}\else $\smash{\calm^\sharp}$\fi}
\def\Mflat{\relax\ifmmode{\calm^\flat}\else $\smash{\calm^\flat}$\fi}
\def\preMcheck{\kern2pt\hbox{\Bigcheck\kern-12pt{$\cal M$}}}
\def\Mcheck{\relax\ifmmode\preMcheck\else $\preMcheck$\fi}
\def\preMhat{\kern2pt\hbox{\Bighat\kern-12pt{$\cal M$}}}
\def\Mhat{\relax\ifmmode\preMhat\else $\preMhat$\fi}

\def\Bsharp{\relax\ifmmode{\calb^\sharp}\else $\calb^\sharp$\fi}
\def\Bflat{\relax\ifmmode{\calb^\flat}\else $\calb^\flat$ \fi}
\def\preBcheck{\hbox{\Bigcheck\kern-9pt{$\cal B$}}}
\def\Bcheck{\relax\ifmmode\preBcheck\else $\preBcheck$\fi}
\def\preBhat{\hbox{\Bighat\kern-9pt{$\cal B$}}}
\def\Bhat{\relax\ifmmode\preBhat\else $\preBhat$\fi}

\def\figBcheck{\kern3pt\hbox{\raise1pt\hbox{\bigboldcheck}\kern-11pt
    {\twelvecal B}}}
\def\figBsharp{{\twelvecal B}\raise5pt\hbox{$\twelvemath\sharp$}}
\def\figBflat{{\twelvecal B}\raise5pt\hbox{$\twelvemath\flat$}}

\def\gcheck{\hbox{\lower2.5pt\hbox{\Bigcheck}\kern-8pt$\g$}}
\def\lhat{\hbox{\raise.5pt\hbox{\Bighat}\kern-8pt$\l$}}

\def\Fcheck{\kern2pt\hbox{\raise1pt\hbox{\Bigcheck}\kern-10pt{$\cal F$}}}
\def\Fhat{\kern2pt\hbox{\raise1pt\hbox{\Bighat}\kern-10pt{$\cal F$}}}
 
\def\cp#1{\relax\ifmmode {\IP\kern-2pt{}_{#1}}\else $\IP\kern-2pt{}_{#1}$\fi}
\def\h#1#2{\relax\ifmmode {b_{#1#2}}\else $b_{#1#2}$\fi}

\def\half{{1\over 2}}

\def\frac#1#2{{#1\over #2}}

\def\pd#1#2{{\partial #1\over\partial #2}}

\def\cone{\relax\thinspace\hbox{$<\kern-.8em{)}$}}
\mathchardef\mho"0A30

\def\-{\hphantom{-}}


\def\cmp#1{Commun. Math. Phys. {\bf #1}}


\def\picture #1 by #2 (#3){\vbox to #2{\hrule width #1 height 0pt depth 0pt
                                       \vfill\special{picture #3}}}
\def\scaledpicture #1 by #2 (#3 scaled #4){{\dimen0=#1 \dimen1=#2
           \divide\dimen0 by 1000 \multiply\dimen0 by #4
            \divide\dimen1 by 1000 \multiply\dimen1 by #4
            \picture \dimen0 by \dimen1 (#3 scaled #4)}}
\def\illustration #1 by #2 (#3){\vbox to #2{\hrule width #1 height 0pt depth 0pt
                                       \vfill\special{illustration #3}}}
\def\scaledillustration #1 by #2 (#3 scaled #4){{\dimen0=#1 \dimen1=#2
           \divide\dimen0 by 1000 \multiply\dimen0 by #4
            \divide\dimen1 by 1000 \multiply\dimen1 by #4
            \illustration \dimen0 by \dimen1 (#3 scaled #4)}}


\def\delaOssa{\nobreak\vskip1truein\hbox to\hsize
       {\hskip 4truein Xenia de la Ossa\hfill}}

\def\hoy{\number\day\space de \ifcase\month\or enero\or febrero\or marzo\or
       abril\or mayo\or junio\or julio\or agosto\or septiembre\or octubre\or
       noviembre\or diciembre\fi\space de \number\year}

\def\cropen#1{\crcr\noalign{\vskip #1}}

\newif\ifproofmode
\proofmodefalse

\newif\ifforwardreference
\forwardreferencefalse

\newif\ifchapternumbers
\chapternumbersfalse

\newif\ifcontinuousnumbering
\continuousnumberingfalse

\newif\iffigurechapternumbers
\figurechapternumbersfalse

\newif\ifcontinuousfigurenumbering
\continuousfigurenumberingfalse

\newif\iftablechapternumbers
\tablechapternumbersfalse

\newif\ifcontinuoustablenumbering
\continuoustablenumberingfalse

\font\eqsixrm=cmr6

\def\marginstyle{\eqsixrm}

\newtoks\chapletter
\newcount\chapno
\newcount\eqlabelno
\newcount\figureno
\newcount\tableno

\chapno=0
\eqlabelno=0
\figureno=0
\tableno=0

\def\chapfolio{\ifnum\chapno>0 \the\chapno\else\the\chapletter\fi}

\def\bumpchapno{\ifnum\chapno>-1 \global\advance\chapno by 1
\else\global\advance\chapno by -1 \setletter\chapno\fi
\ifcontinuousnumbering\else\global\eqlabelno=0 \fi
\ifcontinuousfigurenumbering\else\global\figureno=0 \fi
\ifcontinuoustablenumbering\else\global\tableno=0 \fi}

\def\setletter#1{\ifcase-#1{}\or{}%
\or\global\chapletter={A}%
\or\global\chapletter={B}%
\or\global\chapletter={C}%
\or\global\chapletter={D}%
\or\global\chapletter={E}%
\or\global\chapletter={F}%
\or\global\chapletter={G}%
\or\global\chapletter={H}%
\or\global\chapletter={I}%
\or\global\chapletter={J}%
\or\global\chapletter={K}%
\or\global\chapletter={L}%
\or\global\chapletter={M}%
\or\global\chapletter={N}%
\or\global\chapletter={O}%
\or\global\chapletter={P}%
\or\global\chapletter={Q}%
\or\global\chapletter={R}%
\or\global\chapletter={S}%
\or\global\chapletter={T}%
\or\global\chapletter={U}%
\or\global\chapletter={V}%
\or\global\chapletter={W}%
\or\global\chapletter={X}%
\or\global\chapletter={Y}%
\or\global\chapletter={Z}\fi}

\def\tempsetletter#1{\ifcase-#1{}\or{}%
\or\global\chapletter={A}%
\or\global\chapletter={B}%
\or\global\chapletter={C}%
\or\global\chapletter={D}%
\or\global\chapletter={E}%
\or\global\chapletter={F}%
\or\global\chapletter={G}%
\or\global\chapletter={H}%
\or\global\chapletter={I}%
\or\global\chapletter={J}%
\or\global\chapletter={K}%
\or\global\chapletter={L}%
\or\global\chapletter={M}%
\or\global\chapletter={N}%
\or\global\chapletter={O}%
\or\global\chapletter={P}%
\or\global\chapletter={Q}%
\or\global\chapletter={R}%
\or\global\chapletter={S}%
\or\global\chapletter={T}%
\or\global\chapletter={U}%
\or\global\chapletter={V}%
\or\global\chapletter={W}%
\or\global\chapletter={X}%
\or\global\chapletter={Y}%
\or\global\chapletter={Z}\fi}

\def\chapshow#1{\ifnum#1>0 \relax#1%
\else{\tempsetletter{\number#1}\chapno=#1\chapfolio}\fi}

\def\chaplabel#1{\bumpchapno\ifproofmode\ifforwardreference
\immediate\write1{\noexpand\expandafter\noexpand\def
\noexpand\csname CHAPLABEL#1\endcsname{\the\chapno}}\fi\fi
\global\expandafter\edef\csname CHAPLABEL#1\endcsname
{\the\chapno}\ifproofmode\llap{\hbox{\marginstyle #1\ }}\fi\chapfolio}

\def\chapref#1{\ifundefined{CHAPLABEL#1}??\ifproofmode\ifforwardreference%
\else\write16{ ***Undefined Chapter Reference #1*** }\fi
\else\write16{ ***Undefined Chapter Reference #1*** }\fi
\else\edef\LABxx{\getlabel{CHAPLABEL#1}}\chapshow\LABxx\fi
\ifproofmode\write2{Chapter #1}\fi}

\def\eqnum{\global\advance\eqlabelno by 1
\eqno(\ifchapternumbers\chapfolio.\fi\the\eqlabelno)}

\def\eqlabel#1{\global\advance\eqlabelno by 1 \ifproofmode\ifforwardreference
\immediate\write1{\noexpand\expandafter\noexpand\def
\noexpand\csname EQLABEL#1\endcsname{\the\chapno.\the\eqlabelno?}}\fi\fi
\global\expandafter\edef\csname EQLABEL#1\endcsname
{\the\chapno.\the\eqlabelno?}\eqno(\ifchapternumbers\chapfolio.\fi
\the\eqlabelno)\ifproofmode\rlap{\hbox{\marginstyle #1}}\fi}

\def\eqalignnum{\global\advance\eqlabelno by 1
&(\ifchapternumbers\chapfolio.\fi\the\eqlabelno)}

\def\eqalignlabel#1{\global\advance\eqlabelno by 1 \ifproofmode 
\ifforwardreference\immediate\write1{\noexpand\expandafter\noexpand\def
\noexpand\csname EQLABEL#1\endcsname{\the\chapno.\the\eqlabelno?}}\fi\fi
\global\expandafter\edef\csname EQLABEL#1\endcsname
{\the\chapno.\the\eqlabelno?}&(\ifchapternumbers\chapfolio.\fi
\the\eqlabelno)\ifproofmode\rlap{\hbox{\marginstyle #1}}\fi}

\def\eqref#1{\hbox{(\ifundefined{EQLABEL#1}***)\ifproofmode\ifforwardreference%
\else\write16{ ***Undefined Equation Reference #1*** }\fi
\else\write16{ ***Undefined Equation Reference #1*** }\fi
\else\edef\LABxx{\getlabel{EQLABEL#1}}%
\def\LAByy{\expandafter\stripchap\LABxx}\ifchapternumbers%
\chapshow{\LAByy}.\expandafter\stripeq\LABxx%
\else\ifnum\number\LAByy=\chapno\relax\expandafter\stripeq\LABxx%
\else\chapshow{\LAByy}.\expandafter\stripeq\LABxx\fi\fi)\fi}%
\ifproofmode\write2{Equation #1}\fi}

\def\fignum{\global\advance\figureno by 1
\relax\iffigurechapternumbers\chapfolio.\fi\the\figureno}

\def\figlabel#1{\global\advance\figureno by 1
\relax\ifproofmode\ifforwardreference
\immediate\write1{\noexpand\expandafter\noexpand\def
\noexpand\csname FIGLABEL#1\endcsname{\the\chapno.\the\figureno?}}\fi\fi
\global\expandafter\edef\csname FIGLABEL#1\endcsname
{\the\chapno.\the\figureno?}\iffigurechapternumbers\chapfolio.\fi
\ifproofmode\llap{\hbox{\marginstyle#1
\kern1.2truein}}\relax\fi\the\figureno}

\def\figref#1{\hbox{\ifundefined{FIGLABEL#1}!!!!\ifproofmode\ifforwardreference%
\else\write16{ ***Undefined Figure Reference #1*** }\fi
\else\write16{ ***Undefined Figure Reference #1*** }\fi
\else\edef\LABxx{\getlabel{FIGLABEL#1}}%
\def\LAByy{\expandafter\stripchap\LABxx}\iffigurechapternumbers%
\chapshow{\LAByy}.\expandafter\stripeq\LABxx%
\else\ifnum \number\LAByy=\chapno\relax\expandafter\stripeq\LABxx%
\else\chapshow{\LAByy}.\expandafter\stripeq\LABxx\fi\fi\fi}%
\ifproofmode\write2{Figure #1}\fi}

\def\tabnum{\global\advance\tableno by 1
\relax\iftablechapternumbers\chapfolio.\fi\the\tableno}

\def\tablabel#1{\global\advance\tableno by 1
\relax\ifproofmode\ifforwardreference
\immediate\write1{\noexpand\expandafter\noexpand\def
\noexpand\csname TABLABEL#1\endcsname{\the\chapno.\the\tableno?}}\fi\fi
\global\expandafter\edef\csname TABLABEL#1\endcsname
{\the\chapno.\the\tableno?}\iftablechapternumbers\chapfolio.\fi
\ifproofmode\llap{\hbox{\marginstyle#1
\kern1.2truein}}\relax\fi\the\tableno}

\def\tabref#1{\hbox{\ifundefined{TABLABEL#1}!!!!\ifproofmode\ifforwardreference%
\else\write16{ ***Undefined Table Reference #1*** }\fi
\else\write16{ ***Undefined Table Reference #1*** }\fi
\else\edef\LABtt{\getlabel{TABLABEL#1}}%
\def\LABTT{\expandafter\stripchap\LABtt}\iftablechapternumbers%
\chapshow{\LABTT}.\expandafter\stripeq\LABtt%
\else\ifnum\number\LABTT=\chapno\relax\expandafter\stripeq\LABtt%
\else\chapshow{\LABTT}.\expandafter\stripeq\LABtt\fi\fi\fi}%
\ifproofmode\write2{Table#1}\fi}

\newdimen\sectionskip     \sectionskip=20truept
\newcount\sectno
\def\section#1#2{\sectno=0 \null\vskip\sectionskip
    \centerline{\chaplabel{#1}.~~{\bf#2}}\nobreak\vskip.2truein
    \noindent\ignorespaces}

\def\advancesectno{\global\advance\sectno by 1}
\def\sectfolio{\number\sectno}
\def\subsection#1{\goodbreak\advancesectno\null\vskip10pt
                  \noindent\chapfolio.~\sectfolio.~{\bf #1}
                  \nobreak\vskip.05truein\noindent\ignorespaces}

\def\uttg#1{\null\vskip.1truein
    \ifproofmode \line{\hfill{\bf Draft}:
    UTTG--{#1}--\number\year}\line{\hfill\today}
    \else \line{\hfill UTTG--{#1}--\number\year}
    \line{\hfill\ifcase\month\or January\or February\or March\or April\or May\or June
    \or July\or August\or September\or October\or November\or December\fi
    \space\number\year}\fi}

\def\contents{\noindent
   {\bf Contents\Z}\nobreak\vskip.05truein\noindent\ignorespaces}

\def\getlabel#1{\csname#1\endcsname}
\def\ifundefined#1{\expandafter\ifx\csname#1\endcsname\relax}
\def\stripchap#1.#2?{#1}
\def\stripeq#1.#2?{#2}

%
\catcode`@=11 
\def\space@ver#1{\let\@sf=\empty\ifmmode#1\else\ifhmode%
\edef\@sf{\spacefactor=\the\spacefactor}\unskip${}#1$\relax\fi\fi}
\newcount\referencecount     \referencecount=0
\newif\ifreferenceopen       \newwrite\referencewrite
\newtoks\rw@toks
\def\refmark#1{\relax[#1]}
\def\refend{\refmark{\number\referencecount}}
\newcount\lastrefsbegincount \lastrefsbegincount=0
\def\refsend{\refmark{\count255=\referencecount%
\advance\count255 by -\lastrefsbegincount%
\ifcase\count255 \number\referencecount%
\or\number\lastrefsbegincount,\number\referencecount%
\else\number\lastrefsbegincount-\number\referencecount\fi}}
\def\refch@ck{\chardef\rw@write=\referencewrite
\ifreferenceopen\else\referenceopentrue
\immediate\openout\referencewrite=referenc.texauxil \fi}
%
{\catcode`\^^M=\active 
  \gdef\obeyendofline{\catcode`\^^M\active \let^^M\ }}%
%
{\catcode`\^^M=\active 
  \gdef\ignoreendofline{\catcode`\^^M=5}}
{\obeyendofline\gdef\rw@start#1{\def\t@st{#1}\ifx\t@st\blankend%
\endgroup\@sf\relax\else\ifx\t@st\bl@nkend\endgroup\@sf\relax%
\else\rw@begin#1
\backtotext
\fi\fi}}
{\obeyendofline\gdef\rw@begin#1
{\def\n@xt{#1}\rw@toks={#1}\relax%
\rw@next}}
\def\blankend{}
{\obeylines\gdef\bl@nkend{
}}
\newif\iffirstrefline  \firstreflinetrue
\def\rwr@teswitch{\ifx\n@xt\blankend\let\n@xt=\rw@begin%
\else\iffirstrefline\global\firstreflinefalse%
\immediate\write\rw@write{\noexpand\obeyendofline\the\rw@toks}%
\let\n@xt=\rw@begin%
\else\ifx\n@xt\rw@@d \def\n@xt{\immediate\write\rw@write{%
\noexpand\ignoreendofline}\endgroup\@sf}%
\else\immediate\write\rw@write{\the\rw@toks}%
\let\n@xt=\rw@begin\fi\fi\fi}
\def\rw@next{\rwr@teswitch\n@xt}
\def\rw@@d{\backtotext} \let\rw@end=\relax
\let\backtotext=\relax

\newdimen\refindent     \refindent=30pt
\def\Textindent#1{\noindent\llap{#1\enspace}\ignorespaces}
\def\refitem#1{\par\hangafter=0 \hangindent=\refindent\Textindent{#1}}
\def\REFNUM#1{\space@ver{}\refch@ck\firstreflinetrue%
\global\advance\referencecount by 1 \xdef#1{\the\referencecount}}
\def\refnum#1{\space@ver{}\refch@ck\firstreflinetrue%
\global\advance\referencecount by 1\xdef#1{\the\referencecount}\refend}

\def\REF#1{\REFNUM#1%
\immediate\write\referencewrite{%
\noexpand\refitem{#1.}}%
\begingroup\obeyendofline\rw@start}
\def\ref{\refnum\?%
\immediate\write\referencewrite{\noexpand\refitem{\?.}}%
\begingroup\obeyendofline\rw@start}
\def\Ref#1{\refnum#1%
\immediate\write\referencewrite{\noexpand\refitem{#1.}}%
\begingroup\obeyendofline\rw@start}
\def\REFS#1{\REFNUM#1\global\lastrefsbegincount=\referencecount%
\immediate\write\referencewrite{\noexpand\refitem{#1.}}%
\begingroup\obeyendofline\rw@start}

\def\cite#1{\refmark#1}
\catcode`@=12 
%
%
\input epsf.tex
%
%
\proofmodefalse
\baselineskip=15pt plus 1pt minus 1pt
\parskip=5pt
\chapternumberstrue
\forwardreferencefalse
\figurechapternumberstrue
\tablechapternumberstrue
\ifproofmode
\immediate\openout2=allcrossreferfile \fi
\ifforwardreference\input labelfile
\ifproofmode\immediate\openout1=labelfile \fi\fi
\noblackboxes
\hfuzz=1pt
\vfuzz=2pt
%
%
\def\hourandminute{\count255=\time\divide\count255 by 60
\xdef\hour{\number\count255}
\multiply\count255 by -60\advance\count255 by\time
\hour:\ifnum\count255<10 0\fi\the\count255}

\def\chaplabel#1{\bumpchapno\ifproofmode\ifforwardreference
\immediate\write1{\noexpand\expandafter\noexpand\def
\noexpand\csname CHAPLABEL#1\endcsname{\the\chapno}}\fi\fi
\global\expandafter\edef\csname CHAPLABEL#1\endcsname
{\the\chapno}\ifproofmode
\llap{\hbox{\marginstyle #1\ifnum\chapno > -1\ \else
\hskip1.3truein\fi}}\fi\chapfolio}

\def\section#1#2{\sectno=0 \null\vskip\sectionskip
    \ifnum\chapno > -1 
    \centerline{\fourteenrm\chaplabel{#1}.~~\fourteenbold#2}
    \else
    \centerline{\fourteenbold Appendix\ \chaplabel{#1}: {#2}}\fi
    \nobreak\vskip.2truein
    \noindent\ignorespaces}
\def\subsection#1{\goodbreak\advancesectno\null\vskip10pt
                  \noindent{\it \chapfolio.\sectfolio.~#1}
                  \nobreak\vskip.05truein\noindent\ignorespaces}
\def\subsubsection#1{\goodbreak
                  \noindent$\underline{\hbox{#1}}$
                  \nobreak\vskip-5pt\noindent\ignorespaces}
\def\cite#1{\refmark{#1}}
\def\\{\hfill\break}
\def\cropen#1{\crcr\noalign{\vskip #1}}
\def\contents{\line{{\fourteenbold Contents}\hfill}\nobreak\vskip.05truein\noindent%
              \ignorespaces}

\def\titlebox#1#2{\lower7pt\hbox{%
\hsize=5in\vbox{\vskip5pt\centerline{#1}\vskip5pt\centerline{#2}}}}

\font\mathbb msbm7 at 10pt
\font\sevenmathbb msbm7

\def\Fp{\hbox{\mathbb F}_{\kern-2pt p}}

\def\Fpstar{\Fp^*}

\def\Fq{\hbox{\mathbb F}_q}
\def\sevenFq{\hbox{\sevenmathbb F}_q}
\def\sevenFqstar{\hbox{\sevenmathbb F}^*_q}

\font\bigcmmib=cmmib7 at 14pt

\font\bigcmbsy=cmbsy7 at 14pt

\def\biginfinity{{\bigcmbsy\char'061}}

\def\bb#1{\hbox{\mathbb #1}}
\def\bm{{\bf m}}
\def\bone{{\bf 1}}

\def\notdiv{\hbox{$\not|$\kern3pt}}

\def\ord#1{\ca{O}\kern-2pt\left(#1\right)}

\def\teich{\hbox{Teich}}

\def\E#1#2#3{\ca{E}_{#1#2}\left(#3\right)}
\def\bm{{\bf m}}
%
%
\nopagenumbers\pageno=0
\null\vskip-20pt
\vbox{\baselineskip=12pt
\rightline{\eightrm hep-th/0402133}\vskip-3pt
\rightline{\eightrm \today}
\vskip0.6truein
\centerline{\seventeenrm Calabi-Yau Manifolds}
\vskip.2truein
\centerline{\seventeenrm Over}
\vskip.2truein
\centerline{\seventeenrm Finite Fields, II}
\vskip0.5truein
\centerline{%
      {\csc Philip~Candelas}$^1$,\qquad
      {\csc Xenia~de~la~Ossa}$^1$}}
\vskip.1truein
\centerline{\csc and}
\vskip.1truein
\centerline{\csc \hphantom{$^2$}{\csc Fernando Rodriguez Villegas}$^2$}
\vskip.3truein\bigskip
\centerline{
\vtop{\hsize = 3.0truein
\centerline{$^1$\it Mathematical Institute}
\centerline{\it Oxford University}
\centerline{\it 24-29 St.\ Giles'}
\centerline{\it Oxford OX1 3LB, England}}
\vtop{\hsize = 3.0truein
\centerline{$^2$\it Department of Mathematics}
\centerline{\it University of Texas}
\centerline{\it Austin}
\centerline{\it TX 78712, USA}}}
\vskip0.3in\bigskip
\centerline{\bf ABSTRACT}
\vskip.1truein 
\noindent We study $\z$-functions for a one parameter family of quintic threefolds defined over
finite fields and for their mirror manifolds and comment on their structure. The $\z$-function
for the quintic family involves factors that correspond to a certain pair of genus 4 Riemann
curves. The appearance of these factors is intriguing since we have been unable to `see' these
curves in the geometry of the quintic. Having these $\z$-functions to hand we are led to comment
on their form in the light of mirror symmetry. That some residue of mirror symmetry survives into
the $\z$-functions is suggested by an application of the Weil conjectures to
\cy\ threefolds: the $\z$-functions are rational functions and the degrees of the numerators and
denominators are exchanged between the $\z$-functions for the manifold and its mirror.  It is clear
nevertheless that the $\z$-function, as classically defined, makes an essential distinction between \K\
parameters and the coefficients of the defining polynomial. It is an interesting question whether there
is a `quantum modification' of the $\z$-function that restores the symmetry between the \K\ and complex
structure parameters. We note that the $\z$-function seems to manifest an arithmetic analogue of the large
complex structure limit which involves 5-adic expansion.
\newpage
 %
%
{\baselineskip=10pt 
\contents
\vskip10pt
\item{10.~}Introduction
\vskip10pt
\item{11.~}The $\z$-function for $\ca{M}(\ps)$
\itemitem{\it 11.1~}{\it The Euler curves}
\itemitem{\it 11.2~}{\it The auxilliary curves for $\ps=0,1,\infty$}
\itemitem{\it 11.3~}{\it The $\z$-function for $\ca{A}$ and $\ca{B}$}
\itemitem{\it 11.4~}{\it Form of the $\z$-function}
\vskip10pt
\item{12.~}The $\z$-function for $\ps^5=0,1,\infty$
\itemitem{\it 12.1~}{\it $\ps=0$}
\itemitem{\it 12.2~}{\it $\ps^5=1$}
\itemitem{\it 12.3~}{\it $\ps=\infty$}
\vskip10pt
\item{13.~}The $\z$-function for General $\ps$
\itemitem{\it 13.1~}{\it $\r=4$}
\itemitem{\it 13.2~}{\it $\r=2$}
\itemitem{\it 13.3~}{\it $\r=1$}
\vskip10pt
\item{14.~}The $\z$-function for the mirror manifold
\itemitem{\it 14.1~}{\it Considerations of toric geometry}
\itemitem{\it 14.2~}{\it The $\z$-function for $\ca{W}$}
\vskip10pt
\item{15.~}The 5-adic Expansion of $\z$
\itemitem{\it 15.1~}{\it Results mod 5}
\itemitem{\it 15.2~}{\it Higher terms in the 5-adic expansion}
\vskip10pt
\item{16.~}Open Questions
\vskip10pt 
\item{A.~}The Large Complex Structure Limit
\vskip5pt
\item{B.~}The Frobenius Period
}
\newpage
 %
 %
\headline={\ifproofmode\hfil\eightrm draft:\ \today\
\hourandminute\else\hfil\fi}
\footline={\rm\hfil\folio\hfil}
\pageno=1\chapno=9
\section{intro}{Introduction}
This paper is a continuation of~ 
\Ref{\CDR}{P. Candelas, X. de la Ossa and F. Rodriguez Villegas,\\ 
``Calabi-Yau Manifolds over Finite Fields, I'', hep-th/0012233}\ 
where it was shown that the number of $\Fq$-rational points of the one parameter
family of quintic threefolds $\ca{M}(\ps)$, described by the zero locus of the polynomial
 $$
P(x,\,\ps)~=~\sum_{i=1}^5 x_i^5 - 5 \ps\, x_1 x_2 x_3 x_4 x_5~, \eqlabel{polydef}$$
with $q=p^r$ is a power of a prime $p\neq 5$ and $\ps\in\Fp$, may be computed in terms of the
periods of
$\ca{M}(\ps)$. In the following we shall denote by $N_r(\ps)$ the number of projective solutions
over $\bb{F}_{p^r}$ to the equation $P(x,\,\ps)=0$. 

A first point is that the expressions obtained
in \cite{\CDR} are computable in a practical sense and this allows us to make some observations
on the structure of the $\z$-function for these varieties based on numerical experiment. The
$\z$-function is defined in terms of the $N_r$ by the expression
 $$
\z(t,\ps)~=~\exp\left(\sum_{r=1}^\infty N_r(\ps)\, {t^r\over r}\right)~. \eqlabel{Zdef}$$
and has an interesting structure which we discuss in this article. 

Our principal result in this direction is that for general $\ps$ (that is $\psi^5\neq 0,1,\infty$)
the
$\z$-function has the form
 $$
\z_\ca{M}(t,\ps)~=~
{R_{\bf  1}(t,\ps)\, R_\ca{A}(p^\r t^\r,\ps)^{20\over\r}\, R_\ca{B}(p^\r t^\r,\ps)^{30\over\r}
\over (1-t)(1-pt)(1-p^2t)(1-p^3t)}~.\eqlabel{zetaM}$$
 In this expression the $R$'s are quartic polynomials in their first argument that satisfy the Riemann
relation
 $$
R\left( {1\over p^3 t},\ps \right)~=~{1\over p^6 t^4}\, R(t,\ps) $$
and $\r$ $(=1,2~\hbox{or}~4)$ is the least integer such that $p^\r{-}1$ is divisible by 5.
Thus $R_{\bf  1}$, for example, has the structure
 $$ 
R_{\bf  1}(t,\ps)~=~1 + a_{\bf  1}(\ps)\,t + b_{\bf  1}(\ps)\,pt^2 + a_{\bf  1}(\ps)\,p^3t^3 + p^6 t^4 $$
with $a_{\bf  1}$ and $b_{\bf  1}$ integers that vary with $\ps$. The other factors $R_\ca{A}$ and $R_\ca{B}$
have a similar structure. It is intriguing that these factors are related to certain genus 4
Riemann curves (again for general values of $\ps$) which we call $\ca{A}$ and $\ca{B}$. What is
meant by this is that there is a genus 4 curve $\ca{A}$ with $\z$-function
satisfying
 $$
\z_\ca{A}(u,\ps)~=~{R_\ca{A}(u,\ps)^2\over (1-u)(1-pu)}~,$$
with an analogous relation obtaining for another curve $\ca{B}$.
It is a surprising fact consequent on the simple form of $\z_\ca{M}$ that the entire function is
determined by six coefficients, the two coefficients $a_{\bf  1}(\ps)$ and $b_{\bf  1}(\ps)$ and the analogous
coefficients for $R_\ca{A}$ and $R_\ca{B}$. 

Over $\bb{C}$ the family $\ca{M}(\ps)$ is a one parameter subfamily of a 101-parameter family of
complex manifolds. The number of rational points of $\ca{M}$ over $\Fq$ is in a certain sense a
p-adic period of $\ca{M}$ that varies with the parameters. It is natural when calculating
$N_r(\ps)$ to pass to p-adic analysis and once this is done the computations are in many respects
analogous to processes in the study of the variation of the complex structure of $\ca{M}$. 

Now the variation of complex structure of $\ca{M}(\ps)$ corresponds to a variation of the  
coefficients of its defining polynomial
 $$
P(x,\ps)\longrightarrow P(x,\ps) - 5\d\ps\, x^{\bf 1} + \d c_{\bf v}\,x^{\bf v}~.$$
We shall largely adhere to the conventions and notation of \cite{\CDR}. In particular we employ a multiindex
notation such that $x^{\bf v}=x_1^{v_1}x_2^{v_2}x_3^{v_3}x_4^{v_4}x_5^{v_5}$. We use $\bf 1$ to
denote the vector $(1,1,1,1,1)$ so that $x^{\bf 1}=x_1x_2x_3x_4x_5$ is the fundamental monomial
which, for notational convenience, we denote also by $Q$. There are a total of 126 quintic monomials
however these are to be understood as evaluated modulo the ideal $\ca{I}=\left(\pd{P}{x_i}\right)$,
generated by the partial derivatives of $P$, and this reduces the count to 101. Here we will use
$\bf v$ to label quintic monomials distinct from $Q$. The monomials $Q$ and $x^{\bf v}$ label the
differential equations that the periods satisfy as we see by following the Dwork-Griffiths method
as in
\cite{\CDR}. This amounts to considering all monomials of degree divisible by 5 modulo the ideal
$\ca{I}$. On this set there is a natural operation of multiplication by $Q$, which in the
Dwork-Griffiths method corresponds to differentiating with respect to $\ps$. 
\vskip5pt
 $$
\matrix{1& \longrightarrow& Q& \longrightarrow& Q^2& \longrightarrow& Q^3\cropen{10pt}
         && x^{\bf v}& \longrightarrow& Q\,x^{\bf v}\cr}
\rlap{\hskip1.2in $\matrix{:~\ca{L}_{\bf 1}\cropen{10pt} :~\ca{L}_{\bf v}\cr}$}$$
\vskip5pt
\noindent
Owing to the fact that $Q^4\in \ca{I}$ and $Q^2 x^{\bf v}\in \ca{I}$. This diagram takes care of all
monomials of degree divisible by 5 apart from the degree 10 monomials corresponding to 
${\bf v}=(4,3,2,1,0)$ and its permutations. These are in $\ca{I}$ except when $\ps^5=1$ and
contribute to the $\z$-function for the conifold but not for other values of $\ps$. The upshot is
that the first line leads to a differential equation of degree 4 while the second line gives rise
to a differential equation of degree two for each $\bf v$, that is a hundred differential
equations. The total dimension of the space of solutions is then $204=b^3(\ca{M})$ which is the correct
count for the total number of periods.

Since the number of rational points over $\bb{F}_{p^r}$ may be computed from the periods it follows that
this number admits a decomposition of the form
 $$
N_r(\ps)~=~N_{{\bf 1},r}(\ps) + \sum_{\bf v}N_{{\bf v},r}(\ps) \eqlabel{Ndecomp}$$
with $N_{{\bf v},r}(\ps)$ the contribution from the periods that satisfy the differential
equation with operator $\ca{L}_{\bf v}$. This decomposition is given explicitly by
\eqref{newnewnuq} of
\cite{\CDR} for example.  It is of interest to note that while the $N_r(\ps)$ are of course
positive integers the individual
$N_{{\bf v},r}(\ps)$ are rational but are not in general either positive or integral. 

We shall recall presently the relationship of a \cym\ to its mirror in the context of toric geometry and
reflexive polyhedra, for the case that the manifold can be presented as a hypersurface in a toric
variety. This discussion is applicable to a wider class of examples than the quintic. If, as here, the
weights, $k_i$, of the coordinates divide the degree, that is $k_i|\sum_{j=1}^5 k_j$ for each $i$, then
the mirror manifold may be constructed as the resolution of a quotient $\ca{M}/\G$ of a
subfamily of manifolds $\ca{M}$ that are symmetric under a group of automorphisms $\G$. For the case of
the quintic this is particularly simple. The symmetric manifold is given by quintic as in \eqref{polydef}
and $\G$ is the group of automorphisms
 $$
\G:~~(x_1,\,x_2,\,x_3,\,x_4,\,x_5)\longrightarrow 
(\z^{n_1}x_1,\,\z^{n_2}x_2,\,\z^{n_3}x_3,\,\z^{n_4}x_4,\,\z^{n_5}x_5)$$
with $\z^5=1$ and $\sum_{i=1}^5 n_i \equiv 0 \bmod 5$. One can think of the monomials $x^{\bf v}$, in
these cases, as the characters of the group $\G$ and regard \eqref{Ndecomp} as a decomposition of $N_r$
under the action of $\G$.

The principal reason for being interested in the quintic is that it is a simple example of a \cym\
and that such manifolds enjoy many remarkable properties in virtue of their relation to string
theory and supersymmetry. One of these properties is the existence of a mirror manifold $\ca{W}$
for which the roles of the the complex structure parameters and the \K\ parameters are reversed
with respect to those of $\ca{M}$. It is natural and compelling to enquire how the arithmetic of
$\ca{W}$ is related to that of $\ca{M}$. 

Much is known about the structure of the $\z$-function in virtue of the the Weil conjectures
(since proved). In particular we know even before performing specific computation that the
$\z$-function for $\ca{M}(\ps)$ is a rational funcion of $t$ which has the structure
\vskip-5pt
 $$
\z(t)~=~{\hbox{Numerator of degree $2h^{21}+2$ depending on the complex structure of \ca{M}}\over
\hbox{Denominator of degree $2h^{11}+2$}}~. \eqlabel{Zrat}$$
\vskip5pt\noindent 
It is immediately apparent that the $\z$-function does not treat the complex structure and \K\
parameters symmetrically since the numerator depends nontrivially on the complex structure
parameters, for our family of quintics the numerator varies nontrivially with $\ps$, while on the
other hand the denominator depends only on the number of \K\ parameters.

We would perhaps like to introduce a modified or quantum $\z$-function, $\z^Q$, that respects mirror
symmetry and which would have the property
 $$
\z^Q_\ca{M}(t)~=~{1\over \z^Q_\ca{W}(t)} \eqlabel{Qrel}$$
however such a function cannot be given by the classical definition since this would immediately
contradict \eqref{Zdef} by giving
$N_{\ca{M},r}=-N_{\ca{W},r}$. 
It is of course possible to define
 $$
\z^Q_\ca{M}~=~{\hbox{numerator of $\z_\ca{M}$}\over \hbox{numerator of $\z_\ca{W}$}} $$
which will satisfy the desired relation \eqref{Qrel}; however without a more intrinsic definition
this does not seem to be very fruitful.
While we do not propose here such an intrinsic definition nor do we know that such a definition
exists nevertheless we find it interesting to point out insights from mirror symmetry that
may be pertinent. 

One of these relates to the decomposition of $N_r(\ps)$ into a sum of contributions attributed to
monomials $\bf v$. When, as here, we are concerned with a manifold which is a hypersurface in
a toric variety the mirror relation is particularly clear~
\Ref{\Batyrev}{V. V. Batyrev, Duke math J. 69 (1993) 31.}. 
The monomials $\bf v$ corresponding to
(the deformations of) the defining equation of the hypersurface define a lattice $\L$
together with a lattice polyhedron $\D$. This polyhedron has special properties owing to the fact
that it corresponds to a \cym. In particular it has a unique interior lattice point, corresponding
to the fundamental polynomial $x^{\bf 1}$, and no lattice plane, parallel to a codimension one
face, intervenes between the fundamental monomial and the codimension one faces of $\D$. Such a
polyhedron is said to be reflexive and has a naturally defined dual, $\nabla$, defined in the
lattice dual to $\L$, which is also reflexive. The lattice points of $\nabla$ correspond
to the divisors of
$\ca{M}$. More precisely divisors of
$\ca{M}$ correspond to the lattice points of $\nabla$ that are not interior to the codimension one
faces of~$\nabla$. The points of $\nabla$ that are interior to codimension one faces correspond to
divisors of the toric variety that do not intersect the hypersurface $\ca{M}$, hence do not give
rise to divisors of $\ca{M}$. It can shown that a \cy\ variety can be recovered from a reflexive
pair $(\D,\nabla)$; the mirror variety being defined by the pair $(\nabla,\D)$. In this way we may
think of $\bf v$'s as monomials corresponding to $\ca{M}$ or equivalently as divisors of~$\ca{W}$.
The points of~$\D$ that are interior to codimension one faces are then either the monomials $\bf
v$ of $\ca{M}$ in the ideal $\ca{I}$ or the divisors of the toric variety in which $\ca{W}$ embeds
that do not intersect $\ca{W}$. These observations suggest that the $N_{r,{\bf v}}(\ps)$
should correspond in some way to the divisors of the mirror $\ca{W}$ though the correspondence
cannot be as simple as counting the rational points of the divisor since it is known that the
numbers of points on a divisor that arises as a blowup of a singularity does not depend on any
parameters. The numbers $N_{r,{\bf v}}(\ps)$ however do depend nontrivially on $\psi$. Moreover as
remarked previously they are not in general either positive or integral. This does not
necessarily run counter to what we expect. We are accustomed with regard to mirror symmetry to
compute quantities that depend on parameters and that reduce to their classical integer values only
in a `large complex structure limit' that certain parameters tend to limiting values. However taking
this further suppose we were to work with the full 101 parameter space of quintics $N_r$ would
depend on 101 parameters 100 of which we can think of as the \K\ parameters of the
blowups required to resolve the fixed points of $\ca{M}/\G$. Now however the $N_r$ no longer have a
simple decomposition as in \eqref{Ndecomp}.

In mirror symmetry an important role is played by the large complex structure limit. In this limit
the classical geometry of the space of \K\ parameters of $\ca{M}$ coincides with the classical
geometry on the space of complex structures of $\ca{W}$. The form of the $\z$ function is bound
up with this limit as we shall see. We have stated that for the quintic the $\z$-function takes
the form \eqref{zetaM} and we will see later that the $\z$-function for the mirror~is
 $$
\z_\ca{W}(t,\ps)~=~{R_{\bf  1}(t,\ps)\over (1-t)(1-pt)^{101}(1-p^2t)^{101}(1-p^3t)}~.
\eqlabel{zetaW}$$  
Now when comparing the forms of
$\z_\ca{M}$ with $\z_\ca{W}$ it seems natural to  identify the denominator of $\z_\ca{M}$ with
the large complex structure limit of $R_{\bf  1}(t,\ps)$, the numerator of $\z_\ca{W}$. Similarly
we would like to identify the denominator of $\z_\ca{W}$ with the large complex structure limit
of the numerator of $\z_\ca{M}$. This requires
 $$
R_{\bf  1}(t,\infty)~=~(1-t)(1-pt)(1-p^2t)(1-p^3t)$$
and
 $$
\z_\ca{M}(t,\infty)~=~{1\over\z_\ca{W}(t,\infty)}~=~(1-pt)^{100}(1-p^2t)^{100}~.\eqlabel{lcsl}$$
Any attempt to define the quintic for $\ps=\infty$ leads to a variety that is highly singular.
Different ways of defining the singular variety seem to lead to different counts of the number of
rational points hence different values for the $\z$-function. 
There does however seem to be a procedure that yields the desired expressions in an appropriate
limit: this consists of expanding the $\z$-functions as 5-adic
numbers. The desired relations are then recovered in lowest order. This is most simply stated for
primes such that $5|p-1$. If $\z_\ca{M}$ and $\z_\ca{W}$ are given by \eqref{zetaM} and
\eqref{zetaW} then we find that \eqref{lcsl} is true in the following sense
 $$
\z_\ca{M}(t,\ps)~\equiv~{1\over\z_\ca{W}(t,\ps)}~\equiv~(1-pt)^{100}(1-p^2t)^{100}~~~~
\bmod{5^2}$$
moreover this congruence holds independent of $\ps$. Although unusual we can draw an analogy with
a relation that is a true consequence of mirror symmetry for the quintic. In this context let
$\ps$ denote the complex structure parameter for the mirror quintic $\ca{W}$ and $t$ denote the
complex \K\ parameter for the quintic $\ca{M}$. Then it is known~
\Ref{\CDGP}{P.~Candelas, X.~de~la~Ossa, P.~Green, and L.~Parkes,\\
 ``A Pair of Calabi--Yau Manifolds as an Exactly Soluble Superconformal Theory'',
Nucl.~Phys.~B{\bf 359}(1991) 21--74.}\
that the quantum corrected yukawa coupling is given by the expression
 $$
y_{ttt}(\ps)~=~5 + \sum_{k=1}^\infty n_k k^3 {q^k\over 1-q^k}$$
with $q=e^{2\p i t}$. The large complex structure limit is the limit $\ps,\,\Im(t)\to\infty$ and
in this limit the exponential terms, which express the quantum corrections, tend to
zero and $y_{ttt}(\ps)\to 5$, the classical value. On the other hand it has been shown~
\Ref{\LianYau}{B. H. Lian and S.-T. Yau, ``Arithmetic Properties of Mirror Map and
Quantum Coupling'',  \cmp{176} 163 (1996), hep-th/9411234.}\
that $5^3 | n_k k^3$ for each $k$ so that we can in fact write
 $$
{y_{ttt}(\ps)\over 5} \equiv 1~~~~ \bmod{5^2}~.$$ 

The layout of this article is the following. We begin by examining the form of the $\z$-function
for $\ca{M}(\ps)$. It follows from the decomposition $\eqref{Ndecomp}$ and the relation of each
contribution $N_{{\bf v},r}$ to a differential equation of hypergeometric type that each 
$N_{{\bf v},r}$ is related to the periods of one of two Riemann curves which we call $\ca{A}$ and
$\ca{B}$. In this way we learn that $\z_\ca{M}$ is related to the $\z$-functions for $\ca{A}$ and
$\ca{B}$ as in \eqref{zetaM}. In \SS\chapref{Zfn} and \SS\chapref{general} we discuss this form
further. The special values $\ps^5=0,1,\infty$ require special attention and the cases $\r=4,2,1$
are also best separated. Given in \SS\chapref{general} are also tables of explicit values for the
the $(a,b)$ coefficients. We turn in \SS\chapref{mirrorquintic} to a description of the mirror
manifold in terms of toric geometry and the Cox variables and to a computation of the
corresponding $\z$-function. Once we have $\z_{\ca{M}}$ and $\z_{\ca{W}}$ to hand we present, in
\SS\chapref{modfive}, what we know of the 5-adic expansion of the $\z$-functions. This expansion
seems to play the role of the large complex structure limit in the arithmetic context.
In \SS\chapref{questions} we set out some open questions chief among these is the elucidation of
the intriguing relation between primes and divisors. Such a relation is of course one of the big
ideas of algebraic number theory. In our context however the p-adic $p$ plays a role analogous to
that of the \K\ form. The fact that $p$ plays a role analogous to the parameter $\ve$ that
appears in the Frobenius period is apparent from \cite{\CDR}. We include in Appendix A a
discussion of the Frobenius period following~
\Ref{\HLY}{S. Hosono, B.H. Lian and S.-T. Yau,\\ 
``GKZ Generalized Hypergeometric Systems in Mirror Symmetry of Calabi-Yau Hypersurfaces'', 
Commun. Math. Phys. {\bf 182} 535-578 (1996), alg-geom/9511001.}\ 
and~
\Ref{\Stienstra}{J. Stienstra, ``Resonant Hypergeometric Systems and Mirror Symmetry'', in the  
Proceedings of the Taniguchi Symposium 1997 ``Integrable Systems and Algebraic  Geometry'', 
alg-geom/9711002.}\
to recall the point that the parameter $\ve$ is in reality the \K\ form. Appendix B shows that
different ways of defining the manifold for $\ps=\infty$ lead to different counts for the
numbers of points. 
\newpage
\font\bigcal cmbsy7 at 14pt
\section{Euler}{The \hbox{\bigcmmib\char'020}-function for
\hbox{{\bigcal\char'115}({\bigcmmib\char'040})}}
Owing to the fact that the $N_r(\ps)$ are additive
in the monomials $\bf v$ the $\z$-function is multiplicative 
 $$
\z_\ca{M}(t,\ps)~=~{R_{\bf 1}(t,\ps)\prod_{\bf v} R_{\bf v}(t,\ps)\over
(1-t)(1-pt)(1-p^2t)(1-p^3t)}$$
Where $R_{\bf 1}$ together with the denominator is the contribution of $N_{{\bf 1},r}$, and
$R_{\bf v}$ is the contribution of $N_{{\bf v},r}$. The product over $\bf v$ runs over monomials
mod $\ca{I}$ of degree 5 except when $\ps^5=1$ when there is a contribution also from monomials
of degree 10. 

In virtue of the Weil conjectures we know that the numerator of $\z_\ca{M}$ is a polynomial of
degree $b^3(\ca{M})=204$. The computation of \SS\chapref{mirrorquintic} reveals that $R_{\bf 1}$ is the
numerator of the $\z$-function for the mirror and is therefore a polynomial of degree 4. A priori
we expect the individual factors $R_{\bf v}$ to be infinite series in $t$. Monomials $\bf v$ that
differ merely by a permutation of their components contribute equally; we are therefore concerned
only with the monomials of Table \tabref{monomials} below. For the conifold we must also take into
account the degree ten terms, that were computed in \SS\chapref{gauss}.3, and which are related
by permutation to
 $$
R_{(4,3,2,1,0)}(t,1)~=~{1\over \big( 1- (p^2 t)^\r\big)^{1\over 5\r}}~.$$

Turning now to the quintic $\bf v$'s it is an observation based on numerical experiment that
although the $R_{\bf v}$ are in general infinite series they combine in pairs to give (fractional
powers of) polynomials. We find, for general $\ps$, that
$$\eqalign{
R_{(4,1,0,0,0)}(t,\ps) \, R_{(3,2,0,0,0)}(t,\ps)~&=~R_\ca{A}(p^\r t^\r,\ps)^{1\over\r}
\cropen{5pt} 
R_{(3,1,1,0,0)}(t,\ps) \, R_{(2,2,1,0,0)}(t,\ps)~&=~R_\ca{B}(p^\r t^\r,\ps)^{1\over\r}\cr}
\eqlabel{RAB}$$  
where $R_\ca{A}(u,\ps)$ and $R_\ca{B}(u,\ps)$ are quartic polynomials in $u$. Moreover
$R_\ca{A}(u,\ps)$ and $R_\ca{B}(u,\ps)$ are related to the $\z$-functions for certain Riemann
curves $\ca{A}$ and $\ca{B}$. Although we do not understand the geometrical origin of these
auxilliary curves we shall see in the following how they arise in relation to the periods that
determine the $N_r$. 

The auxilliary curves arise through a consideration of the periods. Another way of expressing this fact
is the statement that the Jacobian of $\ca{M}$ contains the Jacobians of the curves $\ca{A}$ and $\ca{B}$
as factors. While mysterious it is frequently the case that the Jacobian of an arithmetic variety
contains the Jacobians of lower dimensional varieties in this way. Although the appearance of these
auxilliary curves is a priori unrelated to the fact that we are here dealing with a \cym\ nevertheless
one might hope that mirror symmetry might shed some light on the appearance of the $\ca{A}$ and $\ca{B}$
curves in the present context. So far however we do not have such an interpretation.
\subsection{The Euler Curves}
Classical analysis gives an expression for the hypergeometric functions, corresponding to the
differential operator $\ca{L}_{\bf v}$, in terms of Euler's integral which is of the form
$$
\int dx\, x^{-\alpha/5} (1-x)^{-\beta/5} (1-x/\psi^5)^{-(1-\b/5)}$$
with $(\alpha,\beta)$ determined by $(a,b,c)$ and hence by $\bf v$. The relation between the
parameters~is
 $$
\a~=~5(1-b)~,~~~\b~=~5(1-a)~,$$
where in these relations we use the fact that $c=a+b$ for all the cases we consider. For the
$(a,b,c)$ that concern us $\a$, $\b$, $5-\a$ and $5-\b$ are positive integers.

If we think of Euler's integral as $\int {dx\over y}$ then we are led to curves
$$
\E\a\b{\ps^5}:~~~y^5 = x^{\alpha} (1-x)^{\beta} (1-x/\psi^5)^{5-\b} \eqlabel{Ecurve}$$
which turn out to have genus 4. It is well known but curious that although the hypergeometric function is
symmetric under interchange of $a$ and $b$ the curve $\ca{E}_{\a\b}$ is not manifestly symmetric and is
exchanged with $\ca{E}_{\b\a}$. We shall show presently that $\ca{E}_{\a\b}$ and $\ca{E}_{\b\a}$
are in fact isomorphic. The values of interest for the parameters are as in the table. Owing to
the skewness of Euler's integral there is a choice in the assignment of $(\a,\b)$ to a monomial
and for the first three monomials the choice is the one given above however for the fourth
monomial it is convenient to make the choice that involves an exchange of $a$ and $b$.
\vskip15pt
\vbox{
$$\vbox{\def\skip{\hskip10pt}
\offinterlineskip\halign{
\vrule\hfil\strut\skip $#$\skip\hfil\vrule height 12pt depth 6pt&\hfil\skip $#$\skip\vrule
&\hfil\skip$#$\skip\vrule\cr
\noalign{\hrule}
\bf v & \alpha& \beta\cr
\noalign{\hrule\vskip3pt\hrule}
(4,1,0,0,0)&2&3\cr\noalign{\hrule}
(3,2,0,0,0)&1&4\cr\noalign{\hrule\vskip3pt\hrule}
(3,1,1,0,0)&2&4\cr\noalign{\hrule}
(2,2,1,0,0)&4&3\cr\noalign{\hrule}
}}
$$
\centerline{Table\tablabel{monomials}{~~The monomials and the corresponding parameters}}
}
\vskip5pt
One can show that the curves corresponding to the first two monomials are isomorphic as are the curves
corresponding to the second pair. The key observation in showing that $\E14{\ps^5}\cong\E23{\ps^5}$ is
that $2(1,4) \equiv (2,3) \bmod5$. We wish to compare the curves $\E14{\ps^5}$ and $\E23{\ps^5}$. The
three points $(x,0)$ with $x=0,1,\ps^5$ correspond trivially as do the single points that each curve has at
infinity; so now take $x\neq 0,1,\ps^5,\infty$. The observation $2(1,4) \equiv (2,3) \bmod5$ leads us to
square the equation for $\E14{\ps^5}$ which we write in the~form
 $$
\left({y^2\over 1-x}\right)^5~=~x^2(1-x)^3\left(1-{x\over\ps^5}\right)^2$$
so by setting $\eta={y^2\over 1-x}$ we see that each point of $\E14{\ps^5}$ maps to a point of
$\E23{\ps^5}$. The inverse relation derives from $3(2,3)\equiv (1,4)\bmod5$. We take the equation for
$\E23{\ps^5}$, cube it and write it in the form
 $$
\left({\eta^3\over x(1-x)\left(1-{x/\ps^5}\right)}\right)^5~=~x(1-x)^4\left(1-{x\over \ps^5}\right)$$
and by setting $y={\eta^3\over x(1-x)\left(1-{x/\ps^5}\right)}$ we see that every point of $\E23{\ps^5}$
maps to a point of~$\E14{\ps^5}$ and that this map is the inverse of the map previously given. In
a precisely analogous way one sees that
$\E12{\ps^5}\cong\E43{\ps^5}$ so that the curves corresponding to the second pair of monomials are also
isomorphic.

If instead of multiplying $(1,4)$ by 2 we multiply by 3 and 4 we show that
 $$
\E14{\ps^5}~\cong~\E23{\ps^5}~\cong~\E32{\ps^5}~\cong~\E41{\ps^5}~. \eqlabel{isomorphisms}$$
Thus all the curves $\E\a{,5-\a}{\ps^5}$, $1\leq\a\leq 4$, are isomorphic. The
two new curves on the right of \eqref{isomorphisms} are the Euler curves related to $\E14{\ps^5}$ and
$\E23{\ps^5}$ by the $a,b$ interchange mentioned previously. 

We wish now to show that
 $$
\E14{\ps^5}~\cong~\E14{{\ps^5\over \ps^5 - 1}}~.\eqlabel{symone}$$
The simplest way to see this seems to be to write the equation for $\E14{\ps^5}$ in the form
 $$
(1-x)\left({y\over 1-x}\right)^5~=~x\left(1-{x\over \ps^5}\right)~.$$
We set $u={y\over 1-x}$ and regard the resulting equation as a quadratic in $x$
 $$
{x^2\over \ps^5} - (u^5 + 1)x + u^5 ~=~0~.$$
In order to complete the square we set $v={2x\over\ps^5} - (u^5+1)$ and we find that the curve can be
written in hyperelliptic form
 $$
v^2~=~u^{10} + 2\left(1 - {2\over\ps^5}\right)u^5 + 1 \eqlabel{hyperelliptic}$$
and one easily sees that the map is an isomorphism. Now it is clear that this new curve is isomorphic to the
curve obtained by the replacement $u\to-u$ and that this replacement is equivalent to 
$\ps^5\to{\ps^5\over \ps^5-1}$. The relation \eqref{symone} can also be established by making
transformations of $x$ that preserve the general form of the Euler curve \eqref{Ecurve}.

Once written in hyperelliptic form as in \eqref{hyperelliptic} it is clear that $\ca{E}_{14}$ has an
automorphism, $\s$, of degree 2:
\hbox{$(u,v) \to (1/u,v/u)$}. From this it follows that the Jacobian of the curve is isogenous to the
square of an elliptic curve. We shall shortly see the effect of this on the $\z$-function of the curve.
If we write this automorphism in terms of the original coordinates it has the form
 $$
\s:~(x,y)~\mapsto~\left( {1-x\over 1-x/\ps^5},\, (1-1/\ps^5)
{x(1-x)\over y(1-x/\ps^5)}\right)~.
\eqlabel{map}$$
Written in this way the map makes sense also as an isomorphism 
 $$
\s:~\E\a\b{\ps^5}\to\E{5-\b}{,5-\a}{\ps^5}$$
which is an automorphism for $\a+\b=5$. 

Now by considering multiples of $(2,4)$ and its transpose $(4,2)$ we see that
 $$\eqalign{
&\E24{\ps^5}~\cong~\E43{\ps^5}~\cong~\E12{\ps^5}~\cong~\E31{\ps^5}\cr
&\E42{\ps^5}~\cong~\E34{\ps^5}~\cong~\E21{\ps^5}~\cong~\E13{\ps^5}\cr}$$
and the isomorphism $\s$ shows, for example, that $\E24{\ps^5}\cong\E13{\ps^5}$ so that all these curves are
isomorphic. If we now make the transformation $x\to \ps^5 x$ in the equation for the Euler curve then we find
that
 $$
\E\a\b{\ps^5}~\cong~\E\a{,5-\b}{1\over\ps^5}\eqlabel{symtwo}$$
and applying this to $\ca{E}_{24}$ yields $\E24{\ps^5}~\cong~\E24{1\over\ps^5}$.

 To summarise let us simplify the notation by writing $\ca{E}_{23}=\ca{A}$ and
$\ca{E}_{24}=\ca{B}$. We have learnt that
 $$\eqalign{
&\E23{\ps^5}~\cong~\E14{\ps^5}~\cong~\ca{A}(\ps^5)~\cong~\ca{A}\left({\ps^5\over\ps^5-1}\right)~,\cropen{5pt}
&\E24{\ps^5}~\cong~\E43{\ps^5}~\cong~\ca{B}(\ps^5)~\cong~\ca{B}\left({1\over\ps^5}\right)~.\cr}
\eqlabel{EFrels}
$$
\subsection{The auxilliary curves for $\ps^5=0,1,\infty$}
By the curve $\ca{E}_{\a\b}(\infty)$ we shall simply mean the curve
 $$
\ca{E}_{\a\b}(\infty)~:~~~y^5~=~x^\a (1-x)^\b $$
and by the curve $\ca{E}_{\a\b}(0)$ we shall mean the curve that results from multiplying \eqref{Ecurve} by
$\ps^5$, setting $\eta= (-1)^{5-\b}\ps y$ and then setting $\ps=0$.
 $$
\ca{E}_{\a\b}(0)~:~~~\eta^5~=~x^{5+\a -\b} (1-x)^\b~. $$ 
With these understandings we have from \eqref{EFrels} that 
$\ca{A}(\infty)\cong\ca{A}(1)$ and $\ca{B}(\infty)\cong\ca{B}(0)$.
The curves $\ca{E}_{\a\b}(1)$ take the simple form
 $$
\ca{E}_{\a\b}(1)~:~~~y^5~=~x^\a (1-x)^5 $$
and since these do not depend on $\b$ we have $\ca{E}_{\a\b}(1)\cong\ca{A}(1)\cong\ca{B}(1)$. It is also
straightforward to see that $\ca{A}(0)\cong\ca{B}(0)$. To see this note that
 $$\eqalign{
\ca{A}(0)\cong\ca{E}_{23}(0)~:~~~& \eta^5~=~x^4 (1-x)^3\cr
\ca{B}(0)\cong\ca{E}_{24}(0)~:~~~& \eta^5~=~x^3 (1-x)^4\cr}$$
and the curves on the right are clearly isomorphic. To summarize: we have established the identities
 $$\eqalign{
\ca{A}(0)~&\cong~\ca{B}(0)~\cong~\ca{B}(\infty)\cr
\ca{A}(1)~&\cong~\ca{B}(1)~\cong~\ca{A}(\infty)~.\cr}\eqlabel{singids}$$
\subsection{Form of the $\z$-functions for $\ca{A}$ and $\ca{B}$}
For $\ps^5\neq 0,1,\infty$ the $\z$-functions for $\ca{A}$ and $\ca{B}$ over $\bb{F}_{p^\r}$ have the form
 $$
\z(u)~=~{R(u)^2\over (1-u)(1-p^\r u)}\eqlabel{ZEuler}$$
with $R$ a quartic
 $$
R(u)~=~1 + a\, u + b\, u^2 + a\, p^\r u^3 +p^{2\r} u^4\eqlabel{R}$$
indicating that for these general values of $\ps$ the curves $\ca{A}$ and $\ca{B}$ have genus four. 
The relation that we observe between between the $\z$-function for $\ca{A}$ and the $\z$-function for the \cym,
again for these values of $\ps$, is that the contribution of the pair of monomials $(4,1,0,0,0)$ and
$(3,2,0,0,0)$ corresponding to
$\ca{A}$ is $R_\ca{A}(p^\r t^\r)^{1\over \r}$. The statement for $\ca{B}$ being analogous. We will take
\eqref{ZEuler} to define the polynomials $R_A$ and $R_B$ even for the special values $\ps^5=0,1,\infty$
for which these polynomials are quadrics rather than quartics. Having defined $R_A$ and $R_B$ via
\eqref{ZEuler} we will have to discuss separately the form taken by the $\z$-function for $\M$ in terms
of $R_A$ and $R_B$ for the special values $\ps^5=0,1,\infty$. This we do in the following section.

Now while \eqref{R} is the general form of the function $R$ in many cases the function
factorizes. For example for $\r=4$ and for all the primes we have examined $R$ factorizes, over
$\bb Z$, in the following form
 $$
R(u)~=~(1 + c\, u + p^\r u^2)^2~=~
\big(1 + \k \sqrt{u} + p^{\r\over 2}u\big)^2 \big(1 - \k \sqrt{u} + p^{\r\over 2}u\big)^2~.
$$
While for $\r=2,1$ the function often, but not always, factorizes as $(1 + c\, u + p^\r u^2)^2$.

For the special values $\ps^5=0,1,\infty$ the statements above are modified. The numerator of the
$\z$-function is now of degree four rather than eight, indicating that the curves are now of genus two. We
will continue to write the $\z$-function as in \eqref{ZEuler} but it should be borne in mind that in one
case, when $\ps=0$ and $\r=1$ that $R^2$ is not a square.  
\subsubsection{$\ps~=~0$}
When $\ps=0$ it follows from our observations concerning the Euler curves that \hbox{$R_\ca{A}=R_\ca{B}$}
and for $\r=4,2$ it is an observation based on computer calculations that 
 $$ 
R_\ca{A}(u,0)~=~R_\ca{B}(u,0)~=~\big( 1 + p^{\r\over 2}u\big)^2~. $$
while for $\r=1$ the expression is more complicated since $R^2$ is a quartic that does not
factorize over $\bb{Z}$
 $$
R^2_\ca{A}(u,0)~=~R^2_\ca{B}(u,0)~=~1 + a\, u + b\, u^2 + a\, p u^3 +p^2 u^4
~=~\big( 1 + \a_{+}\, u + p u^2\big)\big( 1 + \a_{-}\, u + p u^2\big)~,$$
the last expression being a factorization over $\bb{Q}[\sqrt{5}]$.
\subsubsection{$\ps^5~=~1$}
In this case we can calculate the number of points rather easily. In virtue of the identities it is
sufficient to consider the curve 
 $$
y^5~=~x(1-x)^5~.$$
For $x\neq 1$ set $\eta=y/(1-x)$ thus $x=\eta^5$ and
 $$
(x,\,y)~=~\big(\eta^5,\, \eta (1-\eta^5)\big)~.$$
The expression on the right yields a point of the curve for every value of $\eta$ however, if $5|q-1$, the
five values of $\eta$ for which $\eta^5=1$ yield the same point, $(1,0)$, of the curve. Thus, counting also
the point at infinity there are $q-3$ points over $\bb{F}_{p^{r\r}}$. Hence
 $$
\z_{\ca{E}(1),p^\r}(u)~=~{(1 - u)^4\over (1-u)(1-p^\r u)}
~~~\hbox{so}~~~R_\ca{A}(u,1)=R_\ca{B}(u,1)~=~(1-u)^2~.$$
\subsubsection{$\ps=\infty$}
In virtue of \eqref{singids} we have
 $$
R_\ca{A}(u,\infty)~=~(1 - u)^2~~~\hbox{and}~~~R_\ca{B}(u,\infty)~=~R_\ca{B}(u,0)~.$$
\subsection{The form of the $\z$-function}
As stated previously we find on the basis of numerical experiment that for general $\ps$ the
$\z$-function satisfies
$$
\z_\ca{M}(t,\ps)~=~
{R_{\bf 1}(t,\ps)\, R_\ca{A}(p^\r t^\r,\ps)^{20\over \r}\, R_\ca{B}(p^\r t^\r,\ps)^{30\over \r}
\over (1-t)(1-pt)(1-p^2t)(1-p^3t)}~.\eqlabel{formzeta}$$

In the following two sections we examine the detailed form of the $\z$-function. The $R$ factors
depend on $\ps$ and $p$ and we examine the various cases in turn.
\newpage
\section{Zfn}{The \hbox{\bigcmmib\char'020}-functions for
\hbox{\bigcmmib\char'040}~=~0,1,\biginfinity} 
\vskip-20pt
\subsection{$\underline{\ps=0 \vrule height0pt depth 3pt width 0pt}$}
Although $\ps=0$ corresponds to a smooth manifold this value requires special treatment owing to the fact
that it is an orbifold point of the moduli space. For $\ps=0$ we have computed the $\z$-function to
$\ord{t^{16}}$ for $p\leq 101$. The form of the
$\z$-function shows a pattern that depends on $\r(p)$.
\subsubsection{$\r=4$ and $\r=2$}
We find 
 $$
R_{\bf  1}(t)~=~\left(1 + p^{3\r\over 2} t^\r\right)^{4\over\r}~~~\hbox{and}~~~
R_{\bf v}(t)~=~\left(1 + p^{3\r\over 2} t^\r\right)^{2\over\r}~~~\hbox{for $\deg({\bf v})=5$}$$
so for these cases we have
 $$
R_{(4,1,0,0,0)}R_{(3,2,0,0,0)}\!=\! R_{(3,1,1,0,0)}R_{(2,2,1,0,0)}\!=\! R_A(p^\r t^\r,0)^{2\over\r}
\!=\! R_B(p^\r t^\r,0)^{2\over\r}\!=\! \left(1 + p^{3\r\over 2} t^\r\right)^{4\over\r}.$$
In these relations $R_A$ and $R_B$ appear squared relative to role for general $\ps$. Thus for these
primes we have
 $$
\z(t,0)~=~{\left(1 + p^{3\r\over 2} t^\r\right)^{204\over\r}\over (1-t)(1-pt)(1-p^2t)(1-p^3t)}~.$$
\subsubsection{$\r=1$}
The structure in this case is more complicated with $R_{\bf  1}$ a polynomial of degree four of
the~form
 $$
R_{\bf  1}~=~1 + a_{\bf  1}\, t + b_{\bf  1}\, pt^2 + a_{\bf  1}\, p^3t^3 + p^6t^4$$
and the degree 5 terms are series that are equal in pairs
 $$
R_{(4,1,0,0,0)}~=~R_{(3,1,1,0,0)}~~~\hbox{and}~~~
R_{(3,2,0,0,0)}~=~R_{(2,2,1,0,0)}~.\eqlabel{paireqs}$$
So in this case also we have
 $$
R_{(4,1,0,0,0)}R_{(3,2,0,0,0)}= R_{(3,1,1,0,0)}R_{(2,2,1,0,0)}= R_A(pt,0)^2
= R_B(pt,0)^2$$ 
with 
 $$
R_\ca{A}(pt,0)^2~=~R_\ca{B}(pt,0)^2~=~1 + c\,pt + d\,p^2t^2 + c\,p^4t^3 + p^6t^4~.$$
The fact that $R_\ca{A}=R_\ca{B}$ was to be anticipated from \eqref{singids};
note however that \eqref{paireqs} is a stronger statement.

The $\z$-function depends, for these primes, on four coefficients that we record, for the first few
cases, in the following table.
\vskip10pt
$$\vbox{\def\skip{\hskip9pt}
\offinterlineskip\halign{
\vrule\strut\skip $#$\skip\hfil\vrule height 14pt depth 8pt&\hfil\skip $#$\skip\vrule
&\hfil\skip$#$\skip\vrule &\hfil\skip $#$\skip\vrule&\hfil\skip $#$\skip\vrule
&\hfil\skip $#$\skip\vrule&\hfil\skip $#$\skip\vrule\cr
\noalign{\hrule}
& a_{\bf  1}\hfil & b_{\bf  1}\hfil& \a_\pm\hfil & c\hfil & d\hfil& \g_\pm\hfil\cr
\noalign{\hrule\vskip3pt\hrule}
p=11 & -89 &  351 &\half(-89 \pm 25\sqrt{5})&  1 & -9&\half(1 \pm 5\sqrt{5})\cr\noalign{\hrule}
p=31 &-409 & 2641 &\half(-409 \pm 125\sqrt{5})& 11 & 61&\half(11 \pm 5\sqrt{5})\cr\noalign{\hrule}
p=41 & 981 & 9211 &\half(981 \pm 25\sqrt{5})& -9 & 71&\half(-9 \pm 5\sqrt{5})\cr\noalign{\hrule}
p=61 & 1111& 10951 &{11\over 2}(101 \pm 25\sqrt{5})& 1 & 91&\half(1 \pm 5\sqrt{5})\cr\noalign{\hrule}
p=71 & 101 & -8379 &\half(101 \pm 1025\sqrt{5})& -19 & 201&\half(-19 \pm 5\sqrt{5})\cr\noalign{\hrule}
p=101& 271 & 7581 &\half(271 \pm 1025\sqrt{5})& -29 & 381&\half(-29 \pm 5\sqrt{5})\cr\noalign{\hrule}
}}$$
\vskip-5pt
\centerline{Table~\tablabel{psiequalszero}~The coefficients that determine $R_{\bf 1}$ and
$R_\ca{A}=R_\ca{B}$ when $\ps=0$ and $\r=1$}
\vskip30pt
Now we may factor 
 $$
1 + a_{\bf  1}t + b_{\bf  1}pt^2 + a_{\bf  1}p^3t^3 + p^6t^4~=~(1 + \a_{+}t + p^3t^2)(1 + \a_{-}t + p^3t^2)$$
with $\a_{+} + \a_{-}=a_{\bf  1}$ and $\a_{+}\a_{-}=p(b_{\bf  1}-2p^2)$. That is $\a_\pm$ are the roots of the
equation
 $$
\a^2 - a_{\bf  1}\a + p(b_{\bf  1}-2p^2)~=~0~.$$ 
In the same way we may write
 $$
1 + c\,pt + d\,p^2t^2 + c\,p^4t^3 + p^6t^4~=~(1 + \g_{+} pt + p^3t^2)(1 + \g_{-} pt +
p^3t^2)$$ with the $\g_\pm$ roots of the equation
 $$
\g^2 - c\g + (d-2p)~=~0~.$$ 
For the primes of the table we also give the corresponding values of the $\a_\pm$ and the $\g_\pm$. Note
that they take values in $\bb{Q}[\sqrt{5}]$.
\subsection{$\underline{\ps^5=1 \vrule height0pt depth 3pt width 0pt}$}
For $\ps^5=1$ the manifold becomes singular and develops 125 nodes. We refer to such a manifold, which
has ordinary double points as its only singularities as a conifold.
For the conifold the factor $R_{\bf 1}(t)$, which in all other cases is a quartic in
$t$, becomes a cubic which moreover factorizes into a linear factor and a quadratic factor
 $$
R_{\bf  1}(t)~=~(1-\e\, p t)\,(1-a\,t+p^3t^2)$$
where $\e=\left({5\over p}\right)$ and $a$ is the $p$-th coefficient in the $q$-expansion of
the eigenform, $f$, found by Schoen~
\Ref{\Schoen}{C. Schoen, ``On the Geometry of a Special Determinantal
Hypersurface associated to the Mumford-Horrocks Vector Bundle'',
J. Reine Angew. Math. {\bf 364} 85-111 (1986).}\
in relation to a rigid \cym\ obtained by resolving the nodes
of the conifold; the eigenform is the unique cusp form of weight $4$ and level $25$ and may be written in
terms of Dedekind's $\eta$-function:
$$
\eqalign{
f&= \eta(q^5)^4\left[\eta(q)^4 + 5\,\eta(q)^3\eta(q^{25}) + 20\,\eta(q)^2\eta(q^{25})^2 +
25\,\eta(q)\eta(q^{25})^3 + 25\,\eta(q^{25})^4\right]\cropen{3pt}  
&=q + q^2 + 7q^3 - 7q^4 + 7q^6 + 6q^7 - 15q^8 + 22q^9 - 43q^{11} -
49q^{12} - 28q^{13} + 6q^{14}\cr  
&+ 41q^{16} + 91q^{17} + 22q^{18} - 35q^{19} + 42q^{21} - 43q^{22} +
162q^{23} - 105q^{24} - 28q^{26} - 35q^{27} \cr 
&-42q^{28}+ 160q^{29} + 42q^{31} + 161q^{32} - 301q^{33} + 91q^{34} - 154q^{36} -314q^{37} - 35q^{38}\cr 
&- 196q^{39} - 203q^{41} + 42q^{42} + 92q^{43} + 301q^{44} + 162q^{46} + 196q^{47} + 287q^{48} 
 +\cdots~.\cr}
$$
For the conifold each quintic monomial $\bf v$ contributes to $N_r$ when $\r|r$, and not
otherwise; the contribution being $-p^r$. Thus $R_{\bf v}=(1-p^\r t^\r)^{1\over \r}$.
The conifold term, due to the monomials of degree 10,
contributes $p^{2r}$ again when
$\r|r$. Thus the $\z$-function for the conifold takes the form
 $$
\z(t,1)~=~
{(1-\e p t)\,(1-at+p^3t^2)\,\big(1 - (pt)^\r\big)^{100\over \r} \over
(1-t)(1-pt)(1-p^2t)(1-p^3t)\,\big(1 - (p^2t)^\r\big)^{24\over \r} }~.     
$$
It is of interest to understand how this function relates to the $\z$-function for Schoen's
manifold and understand also the role played by the factor that arises from the monomials of
degree 10. We may do this by following the process of passing from the a smooth quintic to
the conifold and from the conifold to the resolution. 

The conifold has 125 nodes corresponding to solutions of the equations 
\hbox{${\partial P(x,1)\over\partial x_i}=0$}. In passing from a smooth quintic to the conifold
125
$S^3$'s are blown down to form these nodes however the radii of these $S^3$'s are not all
independent, in fact only $h^{21}=101$ are, so there are 24 relations hence 24 four-cycles are
created. A heuristic sketch is shown below.
It is again simplest to restrict to primes for which 
$\r=1$ and for these cases the previous expression simplifies to
 $$
\z(t,\ps^5=1)~=~
{(1-a_p\,t+p^3t^2)\,(1-p t)^{100} \over (1-t)(1-p^2t)^{25}(1-p^3t)}$$ 
and we see that the role of the monomials of degree ten is to account for the 24 relations
between the nodes and hence the 24 new cycles that are created.
\vskip5pt
\def\Sthreefig{\vbox{\vskip0pt\hbox{\hskip20pt\epsfxsize=5truein\epsfbox{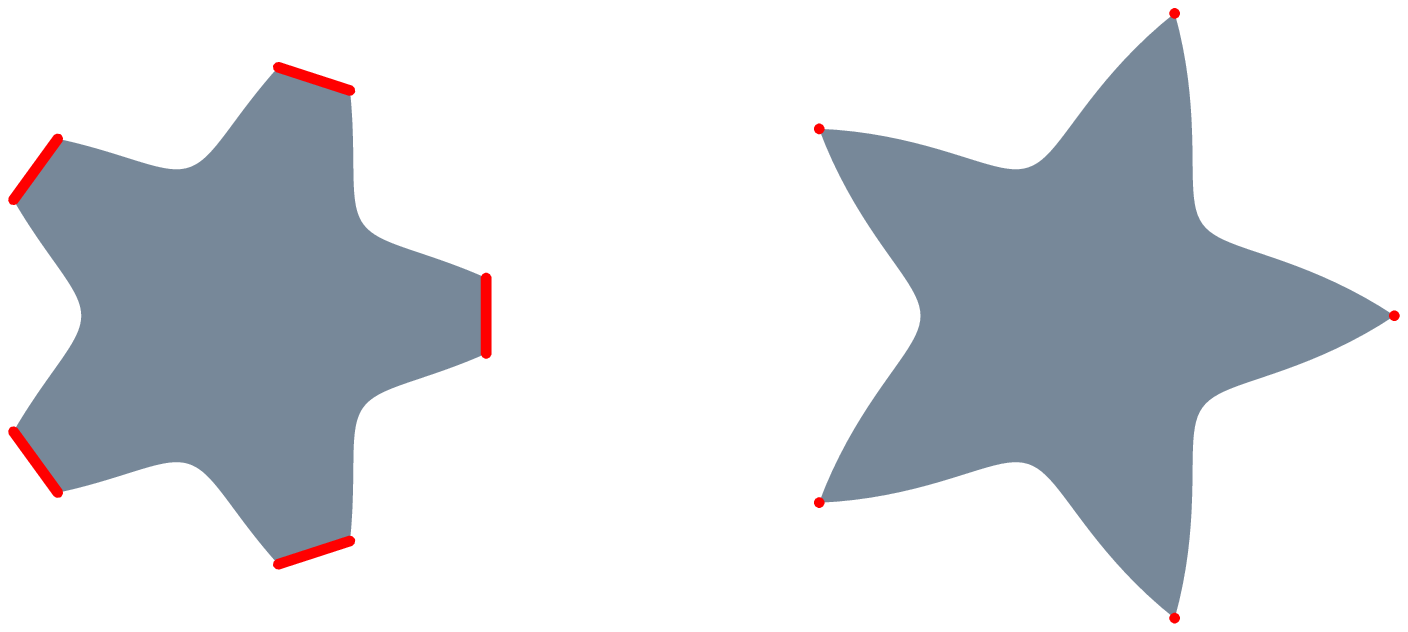}}}}
\figbox{\Sthreefig}{\figlabel{Sthreefig}}{Each relation corresponds to a 4-chain. By shrinking
the boundary of the chain to zero a 4-cycle is created.}
\place{1.1}{2.1}{$24\times$}
\place{3.8}{2.1}{$24\times$}
\place{3.3}{2.05}{$\longrightarrow$}
\place{2.4}{3.0}{$S^3$}
\vskip-5pt
\noindent
If we now resolve 125 nodes with $\bb{P}^1$'s we will have replaced 125 $S^3$'s, each with Euler
number zero, by 125 $S^2$'s each with Euler number 2. Thus the Euler number of the resolved
manifold is $-200+250=50$. Let us denote the resolved manifold by $\widehat\ca{M}$. It is known
that the resolved manifold is rigid, that is $h^{21}(\widehat\ca{M})=0$. Since the Euler number
is 50 we learn that $h^{11}(\widehat\ca{M})=25$. Thus there are 25 parameters corresponding to the
radii of the 125 resolving $S^2$'s so there must be 100 relations and each of these leads to the
destruction of a cycle as sketched in Figure \figref{Stwofig} below. In this way we see that we have
explained why the exponent of the factor $(1-pt)$ in the numerator is $200-100=100$ and that the
$\z$-function for the resolved manifold is
 $$\eqalign{
\z_{\widehat\ca{M}}(t,\ps^5=1)~&=~
{(1-a_p\,t+p^3t^2)\,(1-p t)^{100} \over (1-t)(1-pt)^{125}(1-p^2t)^{25}(1-p^3t)}\cropen{10pt}
&=~
{(1-a_p\,t+p^3t^2) \over (1-t)(1-pt)^{25}(1-p^2t)^{25}(1-p^3t)}\cr}$$
\vskip2pt\noindent
which is the expression obtained by Schoen.
\subsection{$\underline{\ps=\infty \vrule height0pt depth 3pt width 0pt}$}
This is discussed further in Appendix A.
\vskip10pt
\def\Stwofig{\vbox{\vskip0pt\hbox{\hskip20pt\epsfxsize=5truein\epsfbox{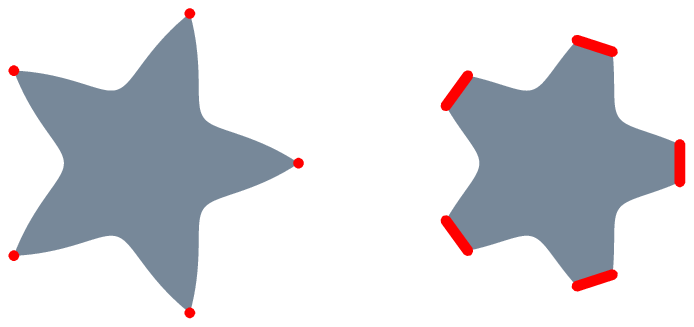}}}}
\figbox{\Stwofig}{\figlabel{Stwofig}}{Here each relation corresponds to a 3-chain. By blowing up
the boundary of the chain to zero a 3-cycle is destroyed.}
\place{1.1}{2.1}{$100\times$}
\place{3.7}{2.1}{$100\times$}
\place{3.3}{2.05}{$\longrightarrow$}
\place{4.95}{3.0}{$S^2$}
\newpage
\section{general}{The \hbox{\bigcmmib\char'020}-function for General
\hskip-2pt\hbox{\bigcmmib\char'040}} 
In all cases the factor $R_{\bf  1}$ is a quartic
 $$
R_{\bf  1}~=~1 + a_{\bf  1} t + b_{\bf  1}pt^2 + a_{\bf  1}p^3t^3 + p^6t^4~.$$
The form of the other factors and the form of the tables shows again a dependence on $\rho(p)$ so we take
these cases in turn.
\subsection{$\underline{\r=4 \vrule height0pt depth 3pt width 0pt}$}
For the quintic $\bf v$'s the factors $R_{\bf v}$ are equal in pairs and take the form
 $$
R_{\bf v}(t)~=~\big( 1 + c_{\bf v}\,p^4t^4 + p^{12}t^8\big)^{1\over 4}~.$$
The values of $(a_{\bf  1},b_{\bf  1})$ and the $c_{\bf v}$ are shown in the following tables for
$p\leq 23$. To compile the tables we have computed $\z$ to $\ord{t^8}$ for $p=3$ and 
$\ord{t^4}$ for $3< p\leq 23$.  In all these cases we find that
$2p^2-c_{\bf v}=\k_{\bf v}^2$ for some integer $\k_{\bf v}$. Thus there is a factorization
 $$
1 + c_{\bf v}p^4t^4 + p^{12}t^8~=~(1 + \k_{\bf v} p^2t^2 + p^6t^4)(1 - \k_{\bf v} p^2t^2 + p^6t^4)~.$$
Note that in all cases $\k_\ca{A}$ is divisible by 10 and $\k_\ca{B}$ is divisible by 5.
\newpage
$$\vbox{\def\skip{\hskip3pt}
\offinterlineskip\halign{
\vrule\strut\hfil\skip $#$\skip\hfil\vrule height 12pt depth 6pt&\hfil\skip $#$\skip\hfil\vrule
&\hfil\skip $#$\skip\hfil\vrule&\hfil\skip $#$\skip\hfil\vrule&\hfil\skip $#$\skip\hfil\vrule
&\hfil\skip $#$\skip\hfil\vrule &\hfil\skip $#$\skip\hfil\vrule&\hfil\skip $#$\skip\hfil\vrule
&\hfil\skip$#$\skip\hfil\vrule\cr
\noalign{\hrule}
p=3 & (a_{\bf  1},b_{\bf  1}) & (4,1,0,0,0) & (3,2,0,0,0) &\k_\ca{A}/10& (3,1,1,0,0) &  (2,2,1,0,0)&\k_\ca{B}/5\cr
\noalign{\hrule\vskip3pt\hrule}
\ps=2 &(5,\, 5.3) &18 &18 &0&- 7  &-7&1 \cr\noalign{\hrule} 
\noalign{\vskip30pt}
\noalign{\hrule}
p=7 & (a_{\bf  1},b_{\bf  1}) & (4,1,0,0,0) & (3,2,0,0,0) &\k_\ca{A}/10& (3,1,1,0,0) &  (2,2,1,0,0)&\k_\ca{B}/5\cr
\noalign{\hrule\vskip3pt\hrule}
\ps=2 &(10,\, 60) &- 2 &- 2 &1& 73 &73 &1 \cr\noalign{\hrule}
\ps=3 &(5,\, 55)  &- 2 &- 2 &1& 98 &98 &0 \cr\noalign{\hrule}
\ps=4 &(-35,\, 115)& 98 &98  &0& 73 &73 &1\cr\noalign{\hrule}
\ps=5 &(-5,\, -30)&- 2 &- 2 &1& 98 &98 &0\cr\noalign{\hrule} 
\ps=6 &(25,\, 50) &- 2 &- 2 &1&- 2 &- 2&2\cr\noalign{\hrule}
\noalign{\vskip30pt}
\noalign{\hrule}
p=13 & (a_{\bf  1},b_{\bf  1}) & (4,1,0,0,0) & (3,2,0,0,0) &\k_\ca{A}/10& (3,1,1,0,0) &  (2,2,1,0,0)&\k_\ca{B}/5\cr
\noalign{\hrule\vskip3pt\hrule}
\ps=2&  (25,\,100)  &238&238&1& 113&113&3  \cr\noalign{\hrule}
\ps=3&  (-25,\,275) &238&238&1& 238&238&2  \cr\noalign{\hrule}
\ps=4&  (-120,\,590)&-62&-62&2& 113&113&3  \cr\noalign{\hrule}
\ps=5&  (-25,\,100) &-62&-62&2&-62&-62 &4  \cr\noalign{\hrule}
\ps=6&  (10,\,-70)  &338&338&0& 113&113&3  \cr\noalign{\hrule}
\ps=7&  (-5,\,260)  &-62&-62&2& 113&113&3  \cr\noalign{\hrule}
\ps=8&  (-5,\,160)  &238&238&1& -62&-62&4  \cr\noalign{\hrule}
\ps=9&  (20,\,-90)  &238&238&1& 238&238&2  \cr\noalign{\hrule}
\ps=10&  (15,\,120) &-62&-62&2& 113&113&3  \cr\noalign{\hrule}
\ps=11&  (85,\,405) &338&338&0& 113&113&3  \cr\noalign{\hrule}
\ps=12&  (-15,\,-20)&338&338&0& 238&238&2  \cr\noalign{\hrule}
}}$$               
\newpage
$$\vbox{\def\skip{\hskip3pt}
\offinterlineskip\halign{
\vrule\strut\hfil\skip $#$\skip\hfil\vrule height 12pt depth 6pt&\hfil\skip $#$\skip\hfil\vrule
&\hfil\skip $#$\skip\hfil\vrule&\hfil\skip $#$\skip\hfil\vrule&\hfil\skip $#$\skip\hfil\vrule
&\hfil\skip $#$\skip\hfil\vrule &\hfil\skip $#$\skip\hfil\vrule&\hfil\skip $#$\skip\hfil\vrule
&\hfil\skip$#$\skip\hfil\vrule\cr
\noalign{\hrule}
p=17 & (a_{\bf  1},b_{\bf  1}) & (4,1,0,0,0) & (3,2,0,0,0) &\k_\ca{A}/10& (3,1,1,0,0) &  (2,2,1,0,0)&\k_\ca{B}/5\cr\noalign{\hrule}
\noalign{\hrule\vskip3pt\hrule}
\ps=2& (-25,\,-25)&578&578&0&  -47& -47&5  \cr\noalign{\hrule}
\ps=3& (-45,\,530)&478&478&1& -322&-322&6  \cr\noalign{\hrule}
\ps=4& (-20,\,130)&178&178&2&  478& 478&2  \cr\noalign{\hrule}
\ps=5& (70,\,270) &478&478&1&  553& 553&1  \cr\noalign{\hrule}
\ps=6& (25,\,225) &178&178&2& -322&-322&6  \cr\noalign{\hrule}
\ps=7& (85,\,660) &478&478&1&  553& 553&1  \cr\noalign{\hrule}
\ps=8& (20,\,120) &178&178&2&  178& 178&4  \cr\noalign{\hrule}
\ps=9& (5,\,5)    &478&478&1&  -47& -47&5  \cr\noalign{\hrule}
\ps=10&(-90,\,610)&478&478&1& -322&-322&6  \cr\noalign{\hrule}
\ps=11&(-75,\,150)&478&478&1&  353& 353&3  \cr\noalign{\hrule}
\ps=12&(120,\,570)&478&478&1& -322&-322&6  \cr\noalign{\hrule}
\ps=13& (75,\,100)&478&478&1&  478& 478&2  \cr\noalign{\hrule}
\ps=14& (55,\,280)&578&578&0&  353& 353&3  \cr\noalign{\hrule}
\ps=15&(-140,\,710)&578&578&0& 178& 178&4  \cr\noalign{\hrule}
\ps=16& (15,\,90) &178&178&2&  -47& -47&5  \cr\noalign{\hrule}
}}
$$ 
\newpage
$$\vbox{\def\skip{\hskip3pt}
\offinterlineskip\halign{
\vrule\strut\hfil\skip $#$\skip\hfil\vrule height 12pt depth 6pt&\hfil\skip $#$\skip\hfil\vrule
&\hfil\skip $#$\skip\hfil\vrule&\hfil\skip $#$\skip\hfil\vrule&\hfil\skip $#$\skip\hfil\vrule
&\hfil\skip $#$\skip\hfil\vrule &\hfil\skip $#$\skip\hfil\vrule
&\hfil\skip $#$\skip\hfil\vrule &\hfil\skip $#$\skip\hfil\vrule\cr
\noalign{\hrule}
p=23 & (a_{\bf  1},b_{\bf  1}) & (4,1,0,0,0) & (3,2,0,0,0) &\k_\ca{A}/10& (3,1,1,0,0) &  (2,2,1,0,0)&\k_\ca{B}/5\cr
\noalign{\hrule\vskip3pt\hrule} 
\ps=2&  (-245,\, 1665) &958 & 958&1 &-167&-167&7\cr\noalign{\hrule}
\ps=3&  (-5,\, 210)    &658 & 658&2 &1058&1058&0\cr\noalign{\hrule}
\ps=4&  (50,\, 450)    &158 & 158&3 &158 &158 &6\cr\noalign{\hrule}
\ps=5&  (0,\, 350)     &-542&-542&4 &958 &958 &2\cr\noalign{\hrule}
\ps=6&  (105,\, 365)   &1058&1058&0 &158 &158 &6\cr\noalign{\hrule} 
\ps=7&  (-135,\, 395)  &958 &958 &1 &433 &433 &5\cr\noalign{\hrule}
\ps=8&  (-40,\, -520)  &958 &958 &1 &1058&1058&0\cr\noalign{\hrule}
\ps=9&  (-105,\, 960)  &-542&-542&4 &158 &158 &6\cr\noalign{\hrule}
\ps=10&  (40,\, 670)   &-542&-542&4 &433 &433 &5\cr\noalign{\hrule}
\ps=11&  (160,\, 930)  &1058&1058&0 &958 &958 &2\cr\noalign{\hrule}
\ps=12&  (-100,\, 850) &-542&-542&4 &-167&-167&7\cr\noalign{\hrule}
\ps=13&  (55,\, 865)   &958 &958 &1 &158 &158 &6\cr\noalign{\hrule}
\ps=14&  (130,\, 240)  &658 &658 &2 &958 &958 &2\cr\noalign{\hrule}
\ps=15&  (-115,\, 605) &1058&1058&0 &958 &958 &2\cr\noalign{\hrule}
\ps=16&  (175,\, 1100) &658 &658 &2 &158 &158 &6\cr\noalign{\hrule}
\ps=17&  (105,\, 90)   &958 &958 &1 &833 &833 &3\cr\noalign{\hrule}
\ps=18&  (-15,\, 730)  &658 &658 &2 &158 &158 &6\cr\noalign{\hrule}
\ps=19&  (150,\, 850)  &-542&-542&4 &833 &833 &3\cr\noalign{\hrule}
\ps=20&  (45,\, 335)   &-542&-542&4 &958 &958 &2\cr\noalign{\hrule}
\ps=21&  (-55,\, -90)  & 958& 958&1 &958 &958 &2\cr\noalign{\hrule}
\ps=22&  (-60,\, 220)  & 158& 158&3 &958 &958 &2\cr\noalign{\hrule}  
}}
$$
\newpage
\subsection{$\underline{\r=2 \vrule height0pt depth 3pt width 0pt}$}
In this case the $R_{\bf v}$'s for quintic $\bf v$'s pair together to give a factor whose general form is
$R(p^2t^2)^\half$ with $R(u)=1 + a\, u + b\,u^2 + a\,p^2\, u^3 + p^4\,u^4$.  In
many cases the $R$ factorizes over $\bb{Z}$ in the form $(1+ c\,u + p^2 u^2)^2$. In these cases
the $R_{\bf v}$'s are equal in pairs~and
 $$
R_{\bf v}(t)~=~\big(1 + c_{\bf v}p^2t^2 + p^6t^4\big)^{1\over 2}~.$$
In the table the cases in which two series multiply to a quartic that is not a square is
indicated by a double column entry showing the corresponding values of the coefficients
$(a,b)$ of the quartic. In those cases where
$f$ is a square this is indicated by single column entries giving the value of $c_{\bf v}$. These
computations have been performed to $\ord{t^4}$. 
\vskip10pt
$$\vbox{\def\skip{\hskip5pt}
\offinterlineskip\halign{
\vrule\strut\hfil\skip $#$\skip\hfil\vrule height 12pt depth 6pt&\hfil\skip $#$\skip\hfil\vrule
&\hfil\skip $#$\skip\hfil\vrule&\hfil\skip $#$\skip\hfil\vrule&\hfil\skip $#$\skip\hfil\vrule
&\hfil\skip $#$\skip\hfil\vrule &\hfil\skip $#$\skip\hfil\vrule\cr
\noalign{\hrule}
p=19 & (a_{\bf  1},b_{\bf  1}) & (4,1,0,0,0) & (3,2,0,0,0) & (3,1,1,0,0) &  (2,2,1,0,0)\cr
\noalign{\hrule\vskip3pt\hrule}
\ps=2& (5,\, 222)& 18& 18& \multispan2{\hfil $(41,\,861)$\hfil\vrule }\cr\noalign{\hrule}
\ps=3& (5,\, 197)&\multispan2{\hfil$(16,\,286)$\hfil\vrule }&\multispan2{\hfil
$(61,\,1621)$\hfil\vrule}\cr\noalign{\hrule} 
\ps=4& (-60,\, 122)&\multispan2{\hfil$(61,\,1621)$\hfil\vrule }&\multispan2{\hfil$(41,\,861)$\hfil\vrule
}\cr\noalign{\hrule} 
\ps=5& (45,\, -153)& 18& 18&\multispan2{\hfil$(41,\,861)$\hfil\vrule }\cr\noalign{\hrule} 
\ps=6& (-130,\, 572)&\multispan2{\hfil$(41,\,861)$\hfil\vrule }& -7& -7\cr\noalign{\hrule} 
\ps=7& (15,\, -378)& 18& 18&\multispan2{\hfil $(16,\,286)$\hfil\vrule }\cr\noalign{\hrule} 
\ps=8& (-90,\, 372)&\multispan2{\hfil$(61,\,1621)$\hfil\vrule }&\multispan2{\hfil$(41,\,861)$\hfil\vrule
}\cr\noalign{\hrule} 
\ps=9& (185,\, 1097)&\multispan2{\hfil $(16,\,286)$\hfil\vrule }&\multispan2{\hfil$(61,\,1621)$\hfil\vrule
}\cr\noalign{\hrule} 
\ps=10& (55,\, 597)& 18& 18&\multispan2{\hfil$(41,\,861)$\hfil\vrule }\cr\noalign{\hrule} 
\ps=11& (20,\, 222)&\multispan2{\hfil$(61,\,1621)$\hfil\vrule }&\multispan2{\hfil $(16,\,286)$\hfil\vrule
}\cr\noalign{\hrule} 
\ps=12& (-75,\, 422)&\multispan2{\hfil$(61,\,1621)$\hfil\vrule }&\multispan2{\hfil$(41,\,861)$\hfil\vrule
}\cr\noalign{\hrule} 
\ps=13& (130,\, 822)&\multispan2{\hfil $(16,\,286)$\hfil\vrule }&\multispan2{\hfil$(61,\,1621)$\hfil\vrule
}\cr\noalign{\hrule} 
\ps=14& (20,\, 22)&\multispan2{\hfil$(1,\,-59)$\hfil\vrule }& 13& 13\cr\noalign{\hrule} 
\ps=15& (-110,\, 422)& 38& 38& 13& 13\cr\noalign{\hrule} 
\ps=16& (65,\, 597)&\multispan2{\hfil $(16,\,286)$\hfil\vrule }& -7& -7\cr\noalign{\hrule} 
\ps=17& (25,\, -228)&\multispan2{\hfil$(41,\,861)$\hfil\vrule }&\multispan2{\hfil$(61,\,1621)$\hfil\vrule
}\cr\noalign{\hrule} 
\ps=18& (-120,\, 522)&\multispan2{\hfil$(1,\,-59)$\hfil\vrule }& 13& 13\cr\noalign{\hrule}
}}
$$
\newpage
\vskip10pt
$$\vbox{\def\skip{\hskip5pt}
\offinterlineskip
\halign{\vrule\strut\hfil\skip$#$\skip\hfil\vrule height12pt depth6pt 
&\hfil\skip$#$\skip\hfil\vrule&\hfil\skip$#$\skip\hfil\vrule&\hfil\skip$#$\skip\hfil\vrule
&\hfil\skip$#$\skip\hfil\vrule&\hfil\skip$#$\skip\hfil\vrule&\hfil\skip$#$\skip\hfil\vrule\cr
\noalign{\hrule}
p=29 & (a_{\bf  1},b_{\bf  1}) & (4,1,0,0,0) & (3,2,0,0,0) & (3,1,1,0,0) &  (2,2,1,0,0)\cr
\noalign{\hrule\vskip3pt\hrule}
\psi=2&(-10,\,932)&\multispan2{\hfil$(81,\,3041)$\hfil}\vrule&\multispan2{\hfil$(1,\,1401)$\hfil}\vrule
\cr\noalign{\hrule}
\psi=3&(-105,\,182)&\multispan2{\hfil$(76,\,3126)$\hfil}\vrule&\multispan2{\hfil$(-19,\,241)$\hfil}\vrule
\cr\noalign{\hrule}
\psi=4&(-80,\,-718)&\multispan2{\hfil$(81,\,3041)$\hfil}\vrule&\multispan2{\hfil$(26,\,1851)$\hfil}\vrule
\cr\noalign{\hrule}
\psi=5&(50,\,1132)&\multispan2{\hfil$(1,\,1401)$\hfil}\vrule&\multispan2{\hfil$(56,\,1966)$\hfil}\vrule
\cr\noalign{\hrule}
\psi=6&(240,\,1582)&\multispan2{\hfil$(76,\,3126)$\hfil}\vrule&\multispan2{\hfil$(56,\,1966)$\hfil}\vrule 
\cr\noalign{\hrule}
\psi=7&(-185,\,682)&\multispan2{\hfil$(81,\,3041)$\hfil}\vrule&\multispan2{\hfil$(101,\,4201)$\hfil}\vrule 
\cr\noalign{\hrule}
\psi=8&(-250,\,1482)&\multispan2{\hfil$(81,\,3041)$\hfil}\vrule&\multispan2{\hfil$(101,\,4201)$\hfil}\vrule 
\cr\noalign{\hrule}
\psi=9&(-125,\,107)&\multispan2{\hfil$(56,\,1966)$\hfil}\vrule&\multispan2{\hfil$(26,\,-149)$\hfil}\vrule
\cr\noalign{\hrule}
\psi=10&(45,\,932)&\multispan2{\hfil$(76,\,3126)$\hfil}\vrule&\multispan2{\hfil$(-19,\,241)$\hfil}\vrule
\cr\noalign{\hrule}
\psi=11&(90,\,682)&\multispan2{\hfil$(56,\,1966)$\hfil}\vrule&\multispan2{\hfil$(101,\,4201)$\hfil}\vrule
\cr\noalign{\hrule}
\psi=12&(15,\,-318)&\multispan2{\hfil$(81,\,3041)$\hfil}\vrule&\multispan2{\hfil$(26,\,-149)$\hfil}\vrule
\cr\noalign{\hrule}
\psi=13&(-35,\,432)&\multispan2{\hfil$(-44,\,2166)$\hfil}\vrule&\multispan2{\hfil$(26,\,-149)$\hfil}\vrule
\cr\noalign{\hrule}
\psi=14&(140,\,582)&\multispan2{\hfil$(1,\,1401)$\hfil}\vrule&\multispan2{\hfil$(106,\,4491)$\hfil}\vrule
\cr\noalign{\hrule}
\psi=15&(-175,\,1607)&\multispan2{\hfil$(56,\,1966)$\hfil}\vrule&\multispan2{\hfil$(1,\,1401)$\hfil}\vrule
\cr\noalign{\hrule}
\psi=16&(-15,\,-68)&\multispan2{\hfil$(56,\,1966)$\hfil}\vrule&\multispan2{\hfil$(101,\,4201)$\hfil}\vrule
\cr\noalign{\hrule}
\psi=17&(75,\,1357)&\multispan2{\hfil$(56,\,1966)$\hfil}\vrule&\multispan2{\hfil$(26,\,-149)$\hfil}\vrule
\cr\noalign{\hrule}
\psi=18&(-90,\,-118)&\multispan2{\hfil$(41,\,1321)$\hfil}\vrule&\multispan2{\hfil$(41,\,1321)$\hfil}\vrule
\cr\noalign{\hrule}
\psi=19&(135,\,557)&\multispan2{\hfil$(56,\,1966)$\hfil}\vrule&\multispan2{\hfil$(76,\,3126)$\hfil}\vrule
\cr\noalign{\hrule}
\psi=20&(80,\,782)&\multispan2{\hfil$(56,\,1966)$\hfil}\vrule&\multispan2{\hfil$(101,\,4201)$\hfil}\vrule
\cr\noalign{\hrule}
\psi=21&(-245,\,1732)&\multispan2{\hfil$(116,\,5046)$\hfil}\vrule&\multispan2{\hfil$(41,\,1321)$\hfil}\vrule
\cr\noalign{\hrule}
\psi=22&(420,\,3032)&\multispan2{\hfil$(81,\,3041)$\hfil}\vrule&\multispan2{\hfil$(26,\,1851)$\hfil}\vrule
\cr\noalign{\hrule}
\psi=23&(265,\,1932)&\multispan2{\hfil$(56,\,1966)$\hfil}\vrule&\multispan2{\hfil$(26,\,1851)$\hfil}\vrule
\cr\noalign{\hrule}
\psi=24&(-175,\,1607)&\multispan2{\hfil$(-44,\,2166)$\hfil}\vrule&\multispan2{\hfil$(26,\,1851)$\hfil}\vrule
\cr\noalign{\hrule}
\psi=25&(125,\,232)&\multispan2{\hfil$(81,\,3041)$\hfil}\vrule&\multispan2{\hfil$(101,\,4201)$\hfil}\vrule
\cr\noalign{\hrule}
\psi=26&(-20,\,-718)&\multispan2{\hfil$(81,\,3041)$\hfil}\vrule&\multispan2{\hfil$(76,\,3126)$\hfil}\vrule
\cr\noalign{\hrule}
\psi=27&(-25,\,757)&\multispan2{\hfil$(76,\,3126)$\hfil}\vrule&\multispan2{\hfil$(106,\,4491)$\hfil}\vrule
\cr\noalign{\hrule}
\psi=28&(45,\,482)&\multispan2{\hfil$(41,\,1321)$\hfil}\vrule&\multispan2{\hfil$(116,\,5046)$\hfil}\vrule
\cr\noalign{\hrule} }}$$
\newpage
\subsection{$\underline{\r=1 \vrule height0pt depth 3pt width 0pt}$}
This is similar to the previous case. The factors $R_{\bf v}$ now
multiply in pairs to give a factor $R(pt)=1 + a\,pt + b\,p^2t^2 + a\,p^4t^3 + p^6t^4$. When $R$
does not factorize over $\bb{Z}$ this is indicated by a double column entry showing the
coefficients $(a,b)$. When
$R$ is a square the $R_{\bf v}$'s are equal in pairs and of the form
$R_{\bf v}~=~1 + c_{\bf v}pt + p^3t^2$
and the tables record the values of the coefficients $c_{\bf v}$.
These computations were carried out to
$\ord{t^4}$ for $p=11$ and to $\ord{t^2}$ for $p\geq 31$.
\vskip10pt
$$\vbox{\def\skip{\hskip5pt}
\offinterlineskip\halign{
\vrule\strut\hfil\skip $#$\skip\hfil\vrule height 12pt depth 6pt
&\hfil\skip $#$\skip\hfil\vrule
&\hfil\skip $#$\skip\hfil\vrule&\hfil\skip $#$\skip\hfil\vrule&\hfil\skip $#$\skip\hfil\vrule
&\hfil\skip $#$\skip\hfil\vrule&\hfil\skip $#$\skip\hfil\vrule\cr
\noalign{\hrule}
p=11 & (a_{\bf  1},b_{\bf  1}) & (4,1,0,0,0) & (3,2,0,0,0) & (3,1,1,0,0) &  (2,2,1,0,0)\cr
\noalign{\hrule\vskip3pt\hrule}
\ps^5=10 &(-14,\, 76) &- 2 &- 2 &3 &3 \cr\noalign{\hrule}
\noalign{\vskip30pt}
\noalign{\hrule}
p=31 & (a_{\bf  1},b_{\bf  1}) & (4,1,0,0,0) & (3,2,0,0,0) & (3,1,1,0,0) &  (2,2,1,0,0)\cr
\noalign{\hrule\vskip3pt\hrule}
\psi^5=5&(-109,\,1166)&8&8&3&3\cr\noalign{\hrule}
\psi^5=6&(66,\,566)&-2&-2&\multispan2{\hfil( 1,31)\hfil\vrule}\cr\noalign{\hrule}
\psi^5=25&(391,\,2666)&\multispan2{\hfil (-9,51)\hfil\vrule}&3&3\cr\noalign{\hrule}
\psi^5=26&(66,\,-684)&-2&-2&\multispan2{\hfil (1,31)\hfil\vrule}\cr\noalign{\hrule}
\psi^5=30&(-259,\,1591)&\multispan2{\hfil (1,31)\hfil\vrule}&-2&-2\cr\noalign{\hrule}
\noalign{\vskip30pt}
\noalign{\hrule}
p=41 & (a_{\bf  1},b_{\bf  1}) & (4,1,0,0,0) & (3,2,0,0,0) & (3,1,1,0,0) &  (2,2,1,0,0)\cr
\noalign{\hrule\vskip3pt\hrule}
\psi^5=3&({-269,2336})&\multispan2{\hfil (-9, 71)\hfil\vrule}&\multispan2{\hfil (-9, 71)\hfil\vrule}\cr\noalign{\hrule}
\psi^5=9&({-44, 2886})&\multispan2{\hfil (1, 51)\hfil\vrule}&3&3\cr\noalign{\hrule}
\psi^5=14&({-144, 2086})&8& 8&\multispan2{\hfil (-9, 71)\hfil\vrule}\cr\noalign{\hrule}
\psi^5=27&({-219, 2736})&-2& -2&\multispan2{\hfil (11, 81)\hfil\vrule}\cr\noalign{\hrule}
\psi^5=32&({-44, 386})&\multispan2{\hfil (1,51)\hfil\vrule}&3& 3\cr\noalign{\hrule}
\psi^5=38&({-219, -389})&-2& -2&\multispan2{\hfil (11,  81)\hfil\vrule}\cr\noalign{\hrule}
\psi^5=40&({581, 4761})&\multispan2{\hfil (1, 51)\hfil\vrule}&3& 3\cr\noalign{\hrule}
}}
$$
\newpage
$$\vbox{\def\skip{\hskip5pt}
\offinterlineskip\halign{
\vrule\strut\hfil\skip $#$\skip\hfil\vrule height 12pt depth 6pt
&\hfil\skip $#$\skip\hfil\vrule
&\hfil\skip $#$\skip\hfil\vrule&\hfil\skip $#$\skip\hfil\vrule&\hfil\skip $#$\skip\hfil\vrule
&\hfil\skip $#$\skip\hfil\vrule&\hfil\skip $#$\skip\hfil\vrule\cr
\noalign{\hrule}
p=61 & (a_{\bf  1},b_{\bf  1}) & (4,1,0,0,0) & (3,2,0,0,0) & (3,1,1,0,0) &  (2,2,1,0,0)\cr
\noalign{\hrule\vskip3pt\hrule}
\psi^5=11&(-364,526)&8&8&-7&-7\cr\noalign{\hrule}
\psi^5=13&(-139,-299)&\multispan2{\hfil (1,91)\hfil\vrule}&\multispan2{\hfil (1,91)\hfil\vrule}\cr\noalign{\hrule}
\psi^5=14&(11,-2099)&\multispan2{\hfil (-9,111)\hfil\vrule}&\multispan2{\hfil (11,121)\hfil\vrule}\cr\noalign{\hrule}
\psi^5=21&(636,8526)&\multispan2{\hfil (-9,111)\hfil\vrule}&\multispan2{\hfil (11,121)\hfil\vrule}\cr\noalign{\hrule}
\psi^5=29&(561,8176)&-2&-2&3&3\cr\noalign{\hrule}
\psi^5=32&(-489,3276)&8&8&\multispan2{\hfil (11,121)\hfil\vrule}\cr\noalign{\hrule}
\psi^5=40&(-689,8176)&-2&-2&3&3\cr\noalign{\hrule}
\psi^5=47&(486,326)&\multispan2{\hfil (1,91)\hfil\vrule}&\multispan2{\hfil (1,91)\hfil\vrule}\cr\noalign{\hrule}
\psi^5=48&(11,2276)&\multispan2{\hfil (-9, 111)\hfil\vrule}&\multispan2{\hfil (11,121)\hfil\vrule}\cr\noalign{\hrule}
\psi^5=50&(-489,2651)&\multispan2{\hfil (-9,111)\hfil\vrule}&-7&-7\cr\noalign{\hrule}
\psi^5=60&(-214,3726)&\multispan2{\hfil (6,6)\hfil\vrule}&-2&-2\cr\noalign{\hrule}
\noalign{\vskip30pt}
\noalign{\hrule}
p=71 & (a_{\bf  1},b_{\bf  1}) & (4,1,0,0,0) & (3,2,0,0,0) & (3,1,1,0,0) &  (2,2,1,0,0)\cr
\noalign{\hrule\vskip3pt\hrule}
\psi^5=20&(726, 5996)&\multispan2{\hfil (6, 26)\hfil\vrule}&\multispan2{\hfil (6, 26)\hfil\vrule}\cr\noalign{\hrule}
\psi^5=23&(301, -279)&\multispan2{\hfil (1, 111)\hfil\vrule}&\multispan2{\hfil (11, 141)\hfil\vrule}\cr\noalign{\hrule}
\psi^5=26&(626, 8196)&-2& -2&\multispan2{\hfil (-9, 131)\hfil\vrule}\cr\noalign{\hrule}
\psi^5=30&(-1049, 10046)&8&8&\multispan2{\hfil (21, 221)\hfil\vrule}\cr\noalign{\hrule}
\psi^5=32&(601, 3121)&\multispan2{\hfil (-19, 201)\hfil\vrule}&\multispan2{\hfil (6, 26)\hfil\vrule}\cr\noalign{\hrule}
\psi^5=34&(176, 6846)&-12& -12&\multispan2{\hfil (11, 141)\hfil\vrule}\cr\noalign{\hrule}
\psi^5=37&(-299, 7296)&8& 8&\multispan2{\hfil (-4, 21)\hfil\vrule}\cr\noalign{\hrule}
\psi^5=39&(726, 4746)&\multispan2{\hfil (6, 26)\hfil\vrule}&3& 3\cr\noalign{\hrule}
\psi^5=41&(-624, 1946)&-2& -2&\multispan2{\hfil (-9, 131)\hfil\vrule}\cr\noalign{\hrule}
\psi^5=45&(-549, 1546)&\multispan2{\hfil (-9, 131)\hfil\vrule}&\multispan2{\hfil (21, 221)\hfil\vrule}\cr\noalign{\hrule}
\psi^5=48&(-299, 421)&8&8&\multispan2{\hfil (-4, 21)\hfil\vrule}\cr\noalign{\hrule}
\psi^5=51&(-524, 5996)&\multispan2{\hfil (6, 26)\hfil\vrule}&3&3\cr\noalign{\hrule}
\psi^5=70&(651, 2646)&\multispan2{\hfil (11, 141)\hfil\vrule}&-12&-12\cr\noalign{\hrule}
}}
$$
\newpage
$$\vbox{\def\skip{\hskip5pt}
\offinterlineskip\halign{
\vrule\strut\hfil\skip $#$\skip\hfil\vrule height 12pt depth 6pt
&\hfil\skip $#$\skip\hfil\vrule
&\hfil\skip $#$\skip\hfil\vrule&\hfil\skip $#$\skip\hfil\vrule&\hfil\skip $#$\skip\hfil\vrule
&\hfil\skip $#$\skip\hfil\vrule&\hfil\skip $#$\skip\hfil\vrule\cr
\noalign{\hrule}
p=101 & (a_{\bf  1},b_{\bf  1}) & (4,1,0,0,0) & (3,2,0,0,0) & (3,1,1,0,0) &  (2,2,1,0,0)\cr
\noalign{\hrule\vskip3pt\hrule}
\psi^5= 6 & (-229, -7669) & \multispan2{\hfil (21, 281)\hfil\vrule}& \multispan2{\hfil (-4, 81)\hfil\vrule}\cr\noalign{\hrule}
\psi^5= 10 & (921, 6281) & \multispan2{\hfil (11, 201)\hfil\vrule}& \multispan2{\hfil (6, 86)\hfil\vrule}\cr\noalign{\hrule}
\psi^5= 14 & (-954, 16906) & \multispan2{\hfil (11, 201)\hfil\vrule}& 3& 3 \cr\noalign{\hrule}
\psi^5= 17 & (1521, 14456) & -2& -2 & \multispan2{\hfil (-4, 81)\hfil\vrule}\cr\noalign{\hrule}
\psi^5= 32 & (-929, 6856) & \multispan2{\hfil (1, -79)\hfil\vrule}& \multispan2{\hfil (-9, 191)\hfil\vrule}\cr\noalign{\hrule}
\psi^5= 36 & (-904, 14306) & 8& 8 & \multispan2{\hfil (1, 171)\hfil\vrule}\cr\noalign{\hrule}
\psi^5= 39 & (896, 15706) & -2& -2 & \multispan2{\hfil (-4, 81)\hfil\vrule}\cr\noalign{\hrule}
\psi^5= 41 & (-1504, 19256) & \multispan2{\hfil (6, 86)\hfil\vrule}& \multispan2{\hfil (-14, 126)\hfil\vrule}\cr\noalign{\hrule}
\psi^5= 44 & (-329, -4344) & \multispan2{\hfil (11, 201)\hfil\vrule}& 3& 3 \cr\noalign{\hrule}
\psi^5= 57 & (-229, -4544) & \multispan2{\hfil (21, 281)\hfil\vrule}& \multispan2{\hfil (-4, 81)\hfil\vrule}\cr\noalign{\hrule}
\psi^5= 60 & (196, -11644) & -12& -12 & \multispan2{\hfil (-9, 191)\hfil\vrule}\cr\noalign{\hrule}
\psi^5= 62 & (171, -344) & \multispan2{\hfil (-14, 126)\hfil\vrule}& 3& 3 \cr\noalign{\hrule}
\psi^5= 65 & (2171, 23781) & \multispan2{\hfil (11, 201)\hfil\vrule}& 3& 3 \cr\noalign{\hrule}
\psi^5= 69 & (246, -2994) & \multispan2{\hfil (-19, 261)\hfil\vrule}& \multispan2{\hfil (-14, 126)\hfil\vrule}\cr\noalign{\hrule}
\psi^5= 84 & (1396, 13456) & -2& -2 & \multispan2{\hfil (21, 281)\hfil\vrule}\cr\noalign{\hrule}
\psi^5= 87 & (-279, 556) & 8& 8 & \multispan2{\hfil (1, 171)\hfil\vrule}\cr\noalign{\hrule}
\psi^5= 91 & (796, 3406) & \multispan2{\hfil (-14, 126)\hfil\vrule}& \multispan2{\hfil (6, 86)\hfil\vrule}\cr\noalign{\hrule}
\psi^5= 95 & (-479, -1544) & -2& -2 & \multispan2{\hfil (21, 281)\hfil\vrule}\cr\noalign{\hrule}
\psi^5= 100 & (-1129, 9756) & \multispan2{\hfil (6, 86)\hfil\vrule}& 18& 18 \cr\noalign{\hrule}
}}
$$
\newpage
\section{mirrorquintic}{The \hbox{\bigcmmib\char'020}-function for the Mirror Manifold}
Our aim here is to compute the $\z$-function for the Mirror of the quintic. The most involved part of this
is the computation of the denominator of the $\z$-function which is of degree 204 and encodes the
information of the blow ups that are performed after identifying the points of the quintic under the action
of the group $\G$. 
\subsection{Considerations of toric geometry}
When a \cym\ $\ca{M}$, as is the case for the quintic, can be realised as a hypersurface in a toric variety
then we may associate with $\ca{M}$ its Newton polyhedron, $\D$, which has the property of being reflexive
\cite{\Batyrev}. The deformation class of mirror manifolds may be associated with hypersurfaces in the toric
variety whose fan consists of the cones, with vertex at the unique interior point of $\D$, over the
faces of $\D$. Such a toric manifold will, in general, be singular as will the hypersurfaces since these
will intersect the singular subvarieties of the toric variety. The singularities of the hypersurface are
resolved by resolving the singularities of the ambient toric variety. This is achieved by refining the fan
so that the one dimensional cones of the refined fan are all the lattice points in the faces of $\D$. A
resolution of singularities is thus given by a triangulation of the faces of $\D$. Such a process is, in
general, not unique since there may be many possible triangulations, different triangulations lead to
different resolutions that differ by a sequence of flops. These different manifolds, when they exist,
correspond to the fact that the quantum space of \K\ parameters, which we identify in virtue of mirror
symmetry with the complex structure moduli space of $\ca{M}$, has more than one large complex structure
point. For the mirror quintic there are a great many possible triangulations of the faces but we shall base
our discussion on a maximally symmetric choice. Batyrev\
\Ref{\BatyrevIntegral}{V. Batyrev, ``Birational Calabi-Yau n-folds have Equal Betti Numbers'' 
Proc. European Algebraic  Geometry Conference (Warwick, 1996), alg-geom/9710020}~
has shown by means of $p$-adic integration that the $\z$-function for \cys\ is independent of the
triangulation of the faces of the polyhedron.
 
For the quintic $\D$ is a simplex with vertices corresponding to the monomials $x_i^5$, $i=1,\ldots,5$.
The five facets (top dimensional faces) of $\D$ are three dimensional polyhedra specified by the vertex
$x_i^5$ that does not lie in the facet. Each facet has four two-faces similar to that shown in
Figure~\figref{twoface}. We triangulate the facets by cutting them with the lattice planes that are parallel
to the two faces. This leads to a triangulation of the two faces as indicated in the figure.
\font\sevenbold cmbx7 at 7pt
\vskip0pt
\vbox{
\def\twoface{\vbox{\vskip20pt\hbox{\hskip0pt\epsfxsize=4truein\epsfbox{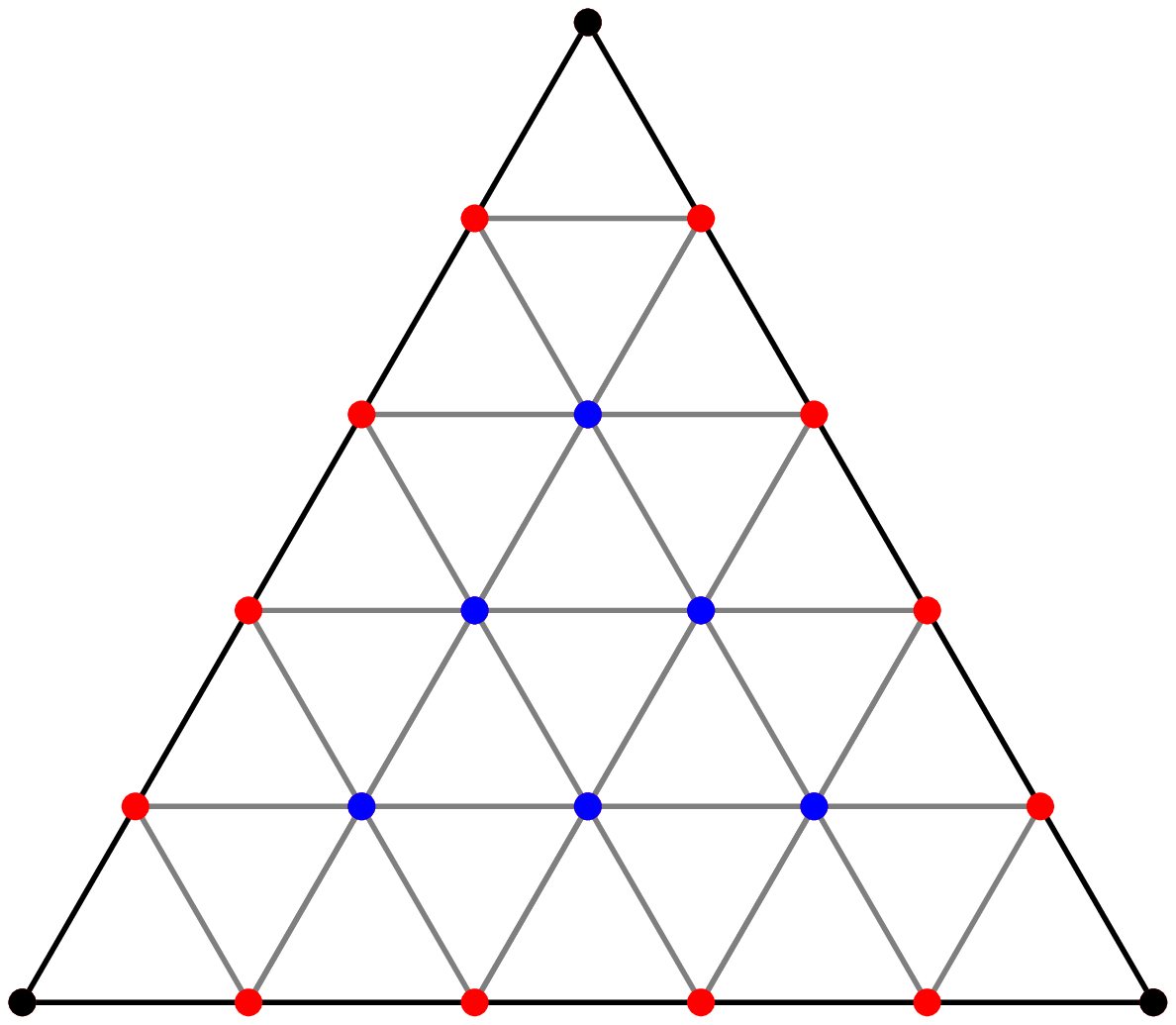}}}}
\figbox{\twoface}{\figlabel{twoface}}{Triangulation of a two-face.}
\place{3.38}{4.2}{\sevenbold (5,0,0,0,0)}
\place{3.73}{3.55}{\sevenbold (4,0,1,0,0)}
\place{4.11}{2.9}{\sevenbold (3,0,2,0,0)}
\place{4.48}{2.25}{\sevenbold (2,0,3,0,0)}
\place{4.86}{1.6}{\sevenbold (1,0,4,0,0)}
\place{5.26}{0.95}{\sevenbold (0,0,5,0,0)}
\place{2.10}{3.55}{\sevenbold (4,1,0,0,0)}
\place{1.75}{2.9}{\sevenbold (3,2,0,0,0)}
\place{1.38}{2.25}{\sevenbold (2,3,0,0,0)}
\place{1}{1.6}{\sevenbold (1,4,0,0,0)}
\place{0.62}{0.95}{\sevenbold (0,5,0,0,0)}
\place{2.94}{3}{\sevenbold (3,1,1,0,0)}
\place{2.55}{2.35}{\sevenbold (2,2,1,0,0)}
\place{3.3}{2.35}{\sevenbold (2,1,2,0,0)}
\place{2.17}{1.7}{\sevenbold (1,3,1,0,0)}
\place{2.93}{1.7}{\sevenbold (1,2,2,0,0)}
\place{3.68}{1.7}{\sevenbold (1,1,3,0,0)}
\place{1.78}{0.75}{\sevenbold (0,4,1,0,0)}
\place{2.55}{0.75}{\sevenbold (0,3,2,0,0)}
\place{3.3}{0.75}{\sevenbold (0,2,3,0,0)}
\place{4.05}{0.75}{\sevenbold (0,1,4,0,0)}
}
We introduce now 125 Cox variables $\tilde{x}_{\bf m}$ for the mirror $\ca{W}$, one for each quintic
monomial
$\bf m$ of $\ca{M}$ apart from the interior point $(1,1,1,1,1)$ which plays a special role. Consider now
all 126 monomials of the polynomial $yP(x)$ which is the quantity that appears as the argument of the
character in the formula $\sum_{y\in\sevenFq} \Th(yP(x))= q\d(P(x))$. To allow for the extra variable $y$
we prepend a 1 to each monomial and stack the monomials $(1,\,{\bf m})$ to form a $126\times 6$ matrix
 $$
M~=~\pmatrix{1&1&1&1&1&1\cr
             1&5&0&0&0&0\cr
             1&0&5&0&0&0\cr
             1&0&0&5&0&0\cr
             1&0&0&0&5&0\cr
             1&0&0&0&0&5\cr
            \vdots&\vdots&\vdots&\vdots&\vdots&\vdots\cr}$$
the ordering is not important as long as we put the interior point first though we find it convenient to list
the five vertices next. Now, by construction, the rows of this matrix give us the monomials of $yP(x)$. The
columns of $M$ give us the monomials of $\tilde{y}\tilde{P}(\tilde{x})$, the analogous polynomial for the
mirror. Let the columns of $M$ be denoted by $(1,{\bf w_j})$, $j=0,\ldots,5$ then the mirror polynomial is
of the form
 $$
\tilde{P}~=~\tilde{x}^{{\bf w}_1} + \tilde{x}^{{\bf w}_2} + \tilde{x}^{{\bf w}_3} + \tilde{x}^{{\bf w}_4}
+\tilde{x}^{{\bf w}_5} -5\tilde{\ps}\, \tilde{x}^{{\bf w}_{\bf  1}}~.\eqlabel{mirrorquintic}$$
The rank of $M$ is 5 so there are 121 vectors $(k_{\bf  1},{\bf k})$ in the null space of $M$ such that 
$(k_{\bf  1},{\bf k})M=0$. Each of these vectors corresponds to a scaling of the coordinates $\tilde x$. We can
choose the scalings such that $\tilde{P}$ is a quintic under one of the scalings and is of degree zero under
the others. A useful basis for this nullspace, which we record here even though we will not need it explicitly,  consists of the vector $(-5;\, {\bf 1};\, {\bf 0})$ and 120 vectors 
$(0;\, {\bf n};\, e_{\bf n})$, where $\bf 1$ denotes the vector $(1,1,1,1,1)$, $\bf n$ runs over the monomials that do not correspond to the vertices of $\D$ and $e_{\bf n}$ is a 120 component vector that has a $1$ in the position corresponding to the row $\bf n$ of $M$ and zeros elsewhere.

The counting is then that we have 125 coordinates $\tilde{x}_{\bf m}$ subject to 121 scalings
and a constraint $\tilde{P}(\tilde{x})=0$ so the resulting manifold has dimension three. It is easy to see
that the polynomial \eqref{mirrorquintic} is, up to redefinitions of the coordinates, the most general
polynomial allowed by the scalings.

The toric variety, $\bb{P}_\D$, is 
 $$
{\bb{C}^{125}\setminus F\over (\bb{C}^*)^{121}}$$
where the role of $F$ is to restrict the $\tilde{x}_{\bf m}$ that can vanish simultaneously. The rule
supplied by toric geometry is that a subset of the coordinates $\tilde{x}_{\bf m}$ can vanish simultaneously
if and only if the corresponding ${\bf m}$'s lie in the same cone. In particular no more than four
coordinates can vanish simultaneously. More specifically we make the following observations:

\noindent 
1) For the purpose of counting the rational points of $\ca{W}$ we need not consider the monomials $\bf m$
interior to the facets since the divisors $\tilde{x}_{(2,1,1,1,0)}=0$ do not intersect $\ca{W}$. To see this
simply note that the coordinate $\tilde{x}_{(2,1,1,1,0)}$ occurs quadratically in $\tilde{x}^{{\bf w}_1}$
and linearly in each of $\tilde{x}^{{\bf w}_2}$, $\tilde{x}^{{\bf w}_3}$ and $\tilde{x}^{{\bf w}_4}$. Thus
when $\tilde{x}_{(2,1,1,1,0)}=0$ the equation $\tilde{P}=0$ reduces to $\tilde{x}^{{\bf w}_5}~=~0$ but the
coordinates that occur in $\tilde{x}^{{\bf w}_5}$ correspond to ${\bf m}$'s whose last component is nonzero
and these do not lie in the same facet as $(2,1,1,1,0)$ and hence cannot lie in the same~cone.

\noindent 
2) To contribute to the count of rational points all the $\tilde{x}_{\bf m}$ that vanish simultaneously must
correspond to $\bf m$'s that lie on the same two-face of $\D$. In other words if three $\bf m$'s lie in a
plane that cuts the interior of the facet then their simultaneous zero locus does not intersect $\ca{W}$. An
example suffices to show why. Take for example the points $(4,1,0,0,0)$, $(4,0,1,0,0)$ and $(4,0,0,1,0)$.
These lie in the same cone however if we set the corresponding coordinates simultaneously to zero the equation
$\tilde{P}=0$ again reduces to
$\tilde{x}^{{\bf w}_5}~=~0$ which has no common solution for $\bf m$ in the same cone.

We have therefore to consider the only the cases of one, two or three $\bf m$'s lying in the same small
triangle of a two face. We shall refer to these cases as points, links and triangles. It is convenient to introduce some specialized notation: rename the first five of the $\tilde{x}$ coordinates $\x_i$,
$i=1,\ldots,5$, that is $\x_1=\tilde{x}_{(5,0,0,0,0)},\x_2=\tilde{x}_{(0,5,0,0,0)},\ldots,\x_5=\tilde{x}_{(0,0,0,0,5)},$ and rename
the remaining coordinates $\eta_{\bf n}$. Take, in the first instance, $5\notdiv q-1$ and consider the cases in turn.
\subsubsection{Points}
Here there are three subcases according as the point is a vertex, lies interior to an edge or lies in the
interior of a two-face. Consider first the case of a vertex say $(5,0,0,0,0)$. We choose a cone containing
this vertex, say the cone that contains also $(4,1,0,0,0)$, $(4,0,1,0,0)$ and $(4,0,0,1,0)$. Setting
$\tilde{x}_{(5,0,0,0,0)}=\x_1=0$ the equation $\tilde{P}=0$ reduces to
 $$
\x_2^5\,\eta^{\hat{\bf w}_2} + \x_3^5\,\eta^{\hat{\bf w}_3} + \x_4^5\,\eta^{\hat{\bf w}_4} + 
\x_5^5\,\eta^{\hat{\bf w}_5} ~=~0~, \eqlabel{mirrorquintic}$$
where the $\hat{\bf w}_j$ are shortened vectors obtained by dropping the first five components of the 
${\bf w}_j$. In the above equation we may assign arbitrary nonzero values to the $\eta_{\bf n}$. It remains
to compute the number of ways in which nonzero values may be assigned to the $\x$'s. Since $5\notdiv q-1$
this is equivalent to asking how many solutions there are with $z_i\neq 0$ to the equation
$z_1+z_2+z_3+z_4=0$. Consider the generalization to $k$ variables of this problem and let the number of
solutions be $r_k$. If we assign arbitrary nonzero values to the variables $z_1,\ldots,z_{k-1}$ and then
solve for $z_k$ then we overcount since the resulting value of $z_k$ might be zero. However by allowing for
this overcounting we see that $r_k=(q-1)^{k-1} - r_{k-1}$ with $r_1=0$. In fact we have
 $$
r_k~=~{(q-1)\over q}\big[(q-1)^{k-1}-(-1)^{k-1}\big]~,$$
an expression that will be useful later.
In this way we see that $r_4=(q-1)(q^2-3q+3)$ and that the number of nonzero solutions to
\eqref{mirrorquintic} is
$(q-1)^{121}(q^2-3q+3)$.

For a point interior to an edge consider for example $(4,1,0,0,0)$. We set $\eta_{(4,1,0,0,0)}=0$ and assign 
arbitrary values to the other $\eta_{\bf n}$. We can now assign arbitrary nonzero values to $\x_1$ and
$\x_2$, and we have to solve
 $$
\x_3^5\,\eta^{\hat{\bf w}_3} + \x_4^5\,\eta^{\hat{\bf w}_4} + \x_5^5\,\eta^{\hat{\bf w}_5} ~=~0~ .$$
For nonzero values of $\x_3$, $\x_4$ and $\x_5$; which has $r_3=(q-1)(q-2)$ solutions. In this way we find
a total of $(q-1)^{122} (q-2)$ nonzero solutions. The answer if we 
choose instead the point $(3,2,0,0,0)$ is the same.

For a point interior to a two-face both choices $(3,1,1,0,0)$ and $(2,2,1,0,0)$ give \hbox{$(q-1)^{123}$}
nonzero solutions.
\topinsert
$$\vbox{\def\skip{\hskip10pt}
\offinterlineskip\halign{
\vrule\hfil\strut\skip #\skip\hfil\vrule height 12pt depth 6pt&\skip #\skip\hfil\vrule
&\hfil\skip$#$\skip\hfil\vrule&\hfil\skip$#$\skip\vrule&\hfil\skip$#$\skip\vrule&\hfil\skip$#$\skip\vrule\cr
\noalign{\hrule}
Number of points & Subsidiary condition& \hbox{Contribution}& z_\x & z_\eta & \d \cr
\noalign{\hrule\vskip3pt\hrule}
Point     & Vertex               & q^2 - 3q + 3 & 1 & 0 & 0\cr
          & Interior to edge     & (q-1)(q-2)   & 0 & 1 & 2\cr
          & Interior to two-face & (q-1)^2      & 0 & 1 & 3\cr
\noalign{\hrule\vskip3pt\hrule}
Link      & Edge link            & q-2        & 1 & 1 & 1\cr
                                 		         &&& 0 & 2 & 2\cr
          & Interior link        & q-1        & 0 & 2 & 3\cr
\noalign{\hrule\vskip3pt\hrule}
Triangle  & Any triangle         & 1          & 1 & 2 & 2\cr
                                 		         &&& 0 & 3 & 3\cr
\noalign{\hrule}
}}$$
\vskip-10pt

\parindent=15pt
{\narrower\noindent Table~\tablabel{contributions}: The contributions to
$\widetilde{\n}_{(0)}/(q-1)^{121}$ of the various combinations of points which have some of their
coordinates equal to zero.\narrower\par}
\endinsert
\subsubsection{Links}
Here the distinction is between a link (with both points in the same cone) with both points lying in an edge,
which we shall refer to as an edge link,
and links with at most one point lying on an edge which we refer to as interior links. An edge link gives
rise to $(q-1)^{121}(q-2)$ nonzero solutions while an interior link gives rise to $(q-1)^{122}$ nonzero
solutions.
\subsubsection{Triangles}
By the foregoing the three points must form a small triangle and it turns out that all small triangles
contribute equally with $(q-1)^{121}$ nonzero solutions.

The contribution of the various cases is summarised in the table. If we introduce some notation we can also
summarise the the results in closed form. Let $z_\x$ and $z_\eta$ denote the number of $\x$ and
$\eta$ coordinates that are simultaneously zero and let $z_\x+\d$ be the number of  monomials of
$\tilde{P}$ that vanish. 
If $\widetilde{\n}_{(0)}$ denotes in this context
the number of rational points with at least one coordinate zero we have
 $$
\widetilde{\n}_{(0)}~=~(q-1)^{120-z_\eta+\d}\,r_{5-z_\x-\d}~=~
{(q-1)^{120-z_\eta+\d}\over q}\big[(q-1)^{4-z_\x-\d}-(-1)^{4-z_\x-\d}\big]\eqlabel{n0}$$
as we can check from the table, though we shall see presently the origin of this expression. 
Summing over the various possibilities we find 
 $$\eqalign{
{\widetilde{\n}_{(0)}\over (q-1)^{121}}~&=~
5(q^2{-}3q{+}3) + 40(q^2{-}3q{+}2) + 60(q^2{-}2q{+}1) +50(q{-}2) + 300(q{-}1) + 250\cr
&=~105\,q^2 + 95\,q + 5~.\cr}\eqlabel{nzero} $$

The surprising fact is that the above expressions for the numbers of points apply equally if $5|q-1$. To see
this we compute $\widetilde{\n}_{(0)}$ in terms of Dwork's character
 $$\eqalign{
q(q-1)^{121}\, \widetilde{\n}_{(0)}~&=~\sum_{\tilde{y}\in\sevenFq}\sum_{\tilde{x}} \Th(\tilde{y}\tilde{P}(\tilde{x}))\cr
&=~(q-1)^{125 - z_\x -z_\eta} + 
\sum_{\tilde{y}\in\sevenFqstar}\sum_{\x_i\atop i\in I}\sum_{\eta_j\atop j\in J}\prod_{i\in I}
\Th(\tilde{y}\,\x_i^5\,\eta^{\widehat{\bf w}_i})\cr
&=~(q-1)^{125 - z_\x -z_\eta} + 
(q-1)^\d\sum_{\tilde{y}\in\sevenFqstar}\sum_{\eta_j\atop j\in J}\prod_{i\in I'}
g_{\bf  1}(\tilde{y}\,\eta^{\widehat{\bf w}_i})\cr
&=~(q-1)^{125 - z_\x -z_\eta} + 
(q-1)^{121 - z_\eta + \d}\hskip-10pt
\sum_{\matrix{\scriptstyle s_i=0,\, i\in I'\cropen{-2pt}  
              \scriptstyle 5|\sum s_i\hfill\cropen{1pt} 
              \scriptstyle 5|\sum s_i{\bf w}_i\cr}}^4
\prod_{i\in I'} G_{-ks_i}~.\cr}\eqlabel{n0eq}
$$
In the first equality the $\tilde{x}$-sum is over values of the coordinates with at least one component
zero. In the second sum the $\widehat{\bf w}_i$ are shortened vectors corresponding to removing the first five
components from the ${\bf w}_i$ and $I$ and $J$ denote the sets of the indices $i$ and $j$ for which the
$\x_i$ and the $\eta_j$ are nonzero. In the third equality $I'$ denotes the set of values for the index $i$
for which the monomial $\tilde{x}^{{\bf w}_i}$ is nonzero. In passing to the third equality we have used
the definition \eqref{gausssums} of $g_{\bf  1}$ and in passing to the last equality we have used the relation
\eqref{yg} and we have summed over $\tilde{y}$ and the $\eta_j$.

We can now show that in the last line of \eqref{n0eq} the only term in the sum over the $G_{-ks_i}$ is the
term with all the $s_i$ vanishing. Thus the final sum is simply $(-1)^{5 - z_\x - \d}$ and this returns us
to \eqref{n0}. To see that the only term in the sum is in fact that corresponding to $s_i=0$ note that we may
as well take the constraints to involve all of the $s_i$ as long as we take $s_i=0$ for $i\not\in I'$. If
$5|\sum s_i{\bf w}_i$ and $5|\sum s_i$ then $5|\sum s_i({\bf w}_i - {\bf w}_{\bf  1})$. The vectors 
${\bf e}_i\define {\bf w}_i - {\bf w}_{\bf  1}$ are just the points of the dual polyhedron $\nabla$ referred to
the interior point as origin. The vectors
${\bf e}_1,{\bf e}_2,{\bf e}_3$ and ${\bf e}_4$ are linearly independent and 
${\bf e}_5= -{\bf e}_1 - {\bf e}_2 - {\bf e}_3 - {\bf e}_4$. It follows that $5|(s_i - s_5)$ but 
$0\leq s_i\leq 4$ so all the $s_i$ are equal and at least one of them is~zero.

Consider now, $\widetilde{N}^*$, the number of solutions modulo the scalings to the equation
\hbox{$\tilde{P}(\tilde{x})=0$} with all coordinates nonzero. Proceeding in a way that is by now familiar 
we~have
 $$
q(q-1)^{121}\, \widetilde{N}^*~=~(q-1)^{125} + 
\sum_{\tilde{y}\in\sevenFqstar}\sum_{\x_i\in\sevenFqstar}\sum_{\eta_j\in\sevenFqstar}
\Th(-5\ps\, \tilde{y}\,\x_1\x_2\x_3\x_4\x_5\,\eta^{\widehat{\bf w}_{\bf  1}})\,
\prod_{i=1}^5 \Th(\tilde{y}\,\x_i^5\,\eta^{\widehat{\bf w}_i})~. \eqlabel{prenstar}$$
If $5\notdiv q-1$ then the $\eta$-sum is trivial since we may simply set all the $\eta$'s to unity and sum
over 120 of the scalings which gives a factor of $(q-1)^{120}$. The remaining sum is one which, apart from
the replacement of $p$ by $q$, we have already seen in \SS\chapref{gauss}.2. The result is
 $$
q(q-1)\, \widetilde{N}^*~=~(q-1)^5 + \sum_{m=0}^{q-2} (-1)^m\, G_{5m}G_{-m}^5\,\teich^m(\l)~.\eqlabel{nstar}$$

If on the other hand $5|q-1$ we set $k=(q-1)/5$, expand the first $\Th$-factor in \eqref{prenstar} and
proceed again as in \SS\chapref{gauss}.2. In this way we find
 $$\eqalign{
q(q-1)^{121}\, \widetilde{N}^*~&=~(q-1)^{125} + 
{1\over q-1}\sum_{\tilde{y}\in\sevenFqstar}\sum_{\eta_j\in\sevenFqstar}\sum_{m=0}^{k-1}
(-1)^m\, \teich^m(\l) G_{5m}\times\cr
&\hskip1truein\prod_{i=1}^5 \teich^{-m}(\tilde{y})\,\teich^{-m\widehat{\bf w}_i}(\eta)\, 
g_{-5m}(\tilde{y}\,\eta^{\widehat{\bf w}_i})\cropen{15pt}
&=~(q-1)^{125} + (q-1)^{120}\sum_{m=0}^{k-1}(-1)^m\,  \teich^m(\l)\, G_{5m}
\hskip-10pt\sum_{\matrix{\scriptstyle s_i=0\cropen{0pt}  
              \scriptstyle 5|\sum s_i\cropen{1pt} 
              \scriptstyle 5|\sum s_i{\bf w}_i\cr}}^4
\hskip-5pt\prod_{i=1}^5 G_{-(m+ks_i)}~.\cr}
$$
We have already discussed the restrictions applying to the last sum, except that now we no longer require
any of the $s_i$ to be zero so the conclusion is now that all the $s_i$ are equal: $s_i=a$, $0\leq a\leq 4$.
The effect of this is to extend the $m$-sum to $0\leq m\leq q-2$ and we recover in this way \eqref{nstar}
which we now see to obtain irrespective of whether $5\notdiv q-1$.

Finally we combine \eqref{nzero} and \eqref{nstar} to find the total number of rational points. Writing 
$\widetilde{N}^*=\widetilde{\n}^*/(q-1)^{121}$ and $\widetilde{N}_{(0)}=\widetilde{\n}_{(0)}/(q-1)^{121}$
we find, after some cancellation, that 
 $$
\widetilde{N}^* + \widetilde{N}_{(0)}~=~ {q^4\over q-1}\sum_{m=0}^{q-2} {G_{5m}\over G_m^5}\,\teich^m(\l) + 
100\,(q+q^2)~. $$
The first expression on the right hand side contributes
 $$
{R_{\bf  1}(t,\ps)\over (1-t)(1-pt)(1-p^2t)(1-p^3t)} $$
to the $\z$-function while $100(q+q^2)$ contributes a factor $(1-pt)^{-100}(1-p^2t)^{-100}$. Hence we have
 $$
\z_\ca{W}(t,\ps)~=~{R_{\bf  1}(t,\ps)\over (1-t)(1-pt)^{101}(1-p^2t)^{101}(1-p^3t)}~.$$

Although we have derived this simple form for the denominator by choosing a very symmetric triangulation the
result holds for any triangulation. This follows from the fact that \eqref{n0} holds for every triangulation
and the entries of Table\tabref{contributions} are determined by $z_\x$, $z_\eta$ and $\d$ and these are
determined by whether, for example, a point is a vertex, interior to an edge or interior to a two face
irrespective of the triangulation. Furthermore any triangulation of a two-face can be obtained by a
succession of moves as in the figure and each such move removes an interior link and adds a new interior
link leaving the number of interior links and triangles unchanged.

\vbox{
\def\flop{\vbox{\vskip20pt\hbox{\hskip0pt\epsfxsize=3truein\epsfbox{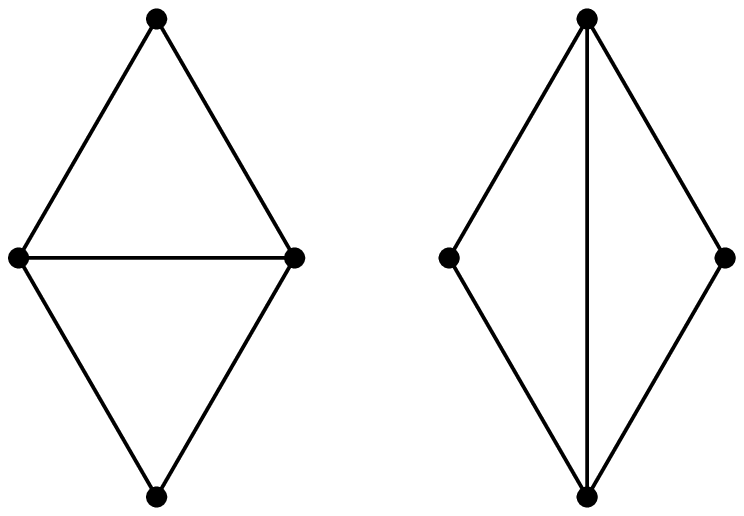}}}}
\figbox{\flop}{\figlabel{flop}}{The fundamental move, corresponding to making a flop, replaces one
diagonal of a quadrilateral by another.} } 
\newpage
\section{modfive}{The 5-adic Expansion of the \hbox{\bigcmmib\char'020}-function}
The results of this section seem to provide an arithmetic analogue of the large complex structure limit
familiar from considerations of mirror symmetry over $\bb{C}$. We are grateful to B.~Gross for suggesting
this line of investigation to us.
\subsection{Results mod 5}
For general $\ps$ and $\ps=0$ the following quantities 
 $$
R_{\bf  1}(t,\ps)~,~~~~R_{\ca{A}}(p^\r t^\r,\ps)^{1\over\r}~,~~~~
R_{\ca{B}}(p^\r t^\r,\ps)^{1\over\r}~,~~~~ (1-t)(1-pt)(1-p^2t)(1-p^3t)$$
are all congruent to $(1-t^\r)^{4\over\r}$ mod 5. For $R_{\bf  1}$, for example, this follows from the
observational fact that, for the primes for which we have computed the coefficients, the coefficients
$(a_{\bf  1},b_{\bf  1})$ are independent of $\ps$ mod 5 and take the values $(0,0)$ for $\r=4$, $(0,2)$ for $\r=2$, and
$(1,1)$ for $\r=1$. Similar facts that are apparent from the tables conspire to make $R_{\ca{A}}$ and
$R_{\ca{B}}$ conform to the stated congruence while for $(1-t)(1-pt)(1-p^2t)(1-p^3t)$ the result is an
elementary consequence of considering the cases $\r=1,2,4$ in turn.

Granted these congruences we have
 $$
\z_{\ca{M}}(t,\ps)\equiv (1-t^\r)^{200\over\r}\equiv (1-t^{25\r})^{8\over\r}\qquad \bmod 5~. $$

Turning to $\z_{\ca{W}}$ one sees that
 $$
\z_{\ca{W}}(t,\ps)\equiv {1\over (1-pt)^{100} (1-p^2t)^{100}}\equiv {1\over (1-t^{25\r})^{8\over\r}}~~~
\hbox{for}~~\r=1,2$$
but the last equality does not hold for $\r=4$. Mirror symmetry seems to be directly applicable
when $5|p-1$ that is when $\r=1$. More generally one should consider $\z$-functions over base
fields which admit nontrivial fifth roots of unity that is $\Fq$ with $p^\r|q$.

Consider now the conifold for which 
 $$
R_{\bf  1}(t,1)~=~(1-\ve_p pt)(1-a_p t + p^3 t^2)$$
and take into account that $a_p\equiv p+p^2 \bmod 5$ (in fact it seems that $a_p\equiv p+p^2 \bmod 25$;
the evidence for this is again observational but based on expanding the modular form to $p_{500}=3571$
terms).

By running through the various values of $p\bmod 5$ we see that
 $$
(1-\ve_p pt)(1-a_p t + p^3 t^2)~\equiv~{(1-t^\r)^{4\over\r}\over 1-t}~$$
so $\z_\ca{M}(t,1)$ differs from the previous value simply by a factor of
$(1-t)^{-1}(1-t^\r)^{-{124\over\r}}$.
\subsection{Higher terms in the 5-adic expansion}
The above congruences can be extended to higher powers of 5 by studying more carefully the structure of the
various coefficients of the tables. In the light of our previous comments regarding the
existence of nontrivial fifth roots of unity we will restrict attention for the remainder of this
section to the case $5|p-1$ and set $k={p-1\over 5}$. We start with the observation that the
quartic
$R_{\bf  1}$ may be written in the form
 $$
R_{\bf  1}(t,\ps)~=~(1-p\,t)^2 (1-p^2 t)^2 + 5^2\,\a(\ps)\, t (1-p\,t) (1-p^2 t) + 5^4\,\b(\ps)\,
p\,t^2~.$$ That a quartic $1+a_{\bf  1} t + b_{\bf  1} p t^2 + a_{\bf  1} p^3 t^3 + p^6 t^4$ can be
written in this form is trivial, the interesting feature however is that the coefficients $\a_{\bf 
1}(\ps)$ and $\b_{\bf  1}(\ps)$ that appear are integers. It is interesting also that this statement is
true if we regard $\ps\in\Fp$ as the fundamental parameter so that the parameter $\l=1/(5\ps)^5$, used
extensively in \cite{\CDR} is necessarily a fifth power. If one defines a manifold as in \eqref{singP} of
Appendix A then one can regard $\l$ as running over all of $\Fp$. In this case weaker congruences obtain
mod 5. Of course the manifolds defined in these two ways are only isomorphic over $\Fp$ when $\l$ is a
fifth power.

A trivial rearrangement of the previous expression yields
 $$
{R_{\bf  1}(t,\ps)\over (1-p\,t)^2 (1-p^2 t)^2}~=~ 
1 + 5^2\,\a_{\bf  1}(\ps)\, {t\over (1-p\,t) (1-p^2 t)} + 
5^4\,\b_{\bf  1}(\ps)\, p\,\left({t\over (1-p\,t) (1-p^2 t)}\right)^2~.
$$
The quadratic factor corresponding to the conifold as well as the denominator of 
$\z_\ca{M}$, which we shall denote by $D(t)$, have a structure similar to that of $R_{\bf  1}$.
 $$\eqalign{
{1-a_p t+p^3\,t^2\over (1-p\,t) (1-p^2 t)}~&=~1 + 5^2\,\a_p {t\over (1-p\,t) (1-p^2 t)}\cropen{10pt}
{D(t)\over (1-p\,t)^2 (1-p^2 t)^2}~&=~1 - (p-1)^2 (p+1)\, {t\over (1-p\,t) (1-p^2 t)}~.\cr}$$
The coefficient $\a_p$ that appears in the first of these relations is an integer furthermore
the coefficient on the right in the second of these expressions is also divisible by~25. The
$\a$'s and $\b$'s also satisfy the congruence 
 $$
\a_\ca{A}(\ps)\equiv - \a_\ca{B}(\ps)\equiv 3\a_{\bf  1}(\ps)+k^2~~~ \bmod 5~.$$ 

The structure revealed by these expansions allows us to to make 5-adic expansions of the polynomials that
make up $\z_\ca{M}$. We find, for example, that the ratio $R_{\bf  1}/D$ has the expansion
 $$
{R_{\bf  1}(t,\ps)\over (1-t) (1-p\,t) (1-p^2t) (1-p^3t)}~=~ \sum_{n=0}^\infty~  
r_n \left({5^2\,t\over (1-p\,t) (1-p^2t)}\right)^n ~~;~~r_0=1~,~~r_n\in\bb{Z} ~.$$
An advantage in organizing the expansion in powers of ${t\over (1-p\,t) (1-p^2 t)}$ is that the
expansion is then invariant under the replacement $t\to {1\over p^3 t}$.

For $R_\ca{A}$ and $R_\ca{B}$ the statements are a little weaker. These polynomials both have expansions of
the form
 $$
{R_\ca{A}(p t,\ps)\over (1-p\,t)^2 (1-p^2 t)^2}~=~1 + 5\,\a_\ca{A}(\ps)\, p\, {t\over (1-p\,t) (1-p^2 t)}
+ 5^2\,\b_\ca{A}(\ps)\, p^2\, \left({t\over (1-p\,t) (1-p^2 t)}\right)^2
 $$ 
with the $\a_\ca{A}(\ps)$ and $\b_\ca{A}(\ps)$ integers. We therefore obtain  $\z_\ca{M}$ mod $5^{n+1}$ by
expanding $\z_\ca{M}/\left((1-p\,t)^{100} (1-p^2 t)^{100}\right)$ as a power series in 
$t/\left((1-p\,t)(1-p^2 t)\right)$ as far as the term of order $n$.

Working mod 125 we find
 $$\eqalign{
{\z_\ca{M}(t,\ps)\over (1-p\,t)^{100} (1-p^2 t)^{100}}~&\equiv~
1 - 25\,\a_\ca{A}(\ps)\, {t\over (1-p\,t) (1-p^2 t)}\cropen{10pt}
(1-p\,t)^{100} (1-p^2 t)^{100}\,\z_\ca{W}(t,\ps)~&\equiv~ 
1 + 50\,\a_\ca{A}(\ps)\, {t\over (1-p\,t) (1-p^2 t)}~.\cr}$$
\newpage
\section{questions}{Open Questions}
Our analysis leaves many questions unanswered. We list here some important loose ends.
\subsection{Primes and \K\ classes}
There is a striking analogy between the hyperplane class, $\ve$, that appears when we use the method of
Fr\"obenius in relation to the periods and the p-adic $p$. The Frobenius period
 $$
\vp(\l,\ve)~=~\l^\ve\sum_{n=0}^\infty a_n(\ve) \,\l^n$$
enters when we calculate the $N_r$ via the periods and semiperiods as in \cite{\CDR}. A first point is
that $\ve$ is in reality the \K\ form of the manifold. This has been made clear by the studies of Hosono,
Lian and Yau \cite{\HLY} and of Stienstra \cite{\Stienstra}. For our situation the relation $\ve^4=0$ is
a reflection of the corresponding relation for the \K\ form. Later when we introduce the first
semiperiod we use instead the relation $\ve^5=0$ corresponding to considering $\ve$ to be the \K\ form for
the embedding $\bb{P}^4$. If there is more that one \K\ parameter then we have $\ve=\sum_k\ve^k e_k$ where
the $\ve^k$ are parameters and the $e_k$ a basis for $H^2$. In particular the $e_k$ satisfy the vanishing
relations appropriate to the intersection ring. Now it is striking also in \cite{\CDR} that the quantity
$\ve$ is analogous to the p-adic $p$. In particular in order to compute the number of rational points we
need only work mod $p^5$. The analogy between primes and divisors is one of the basic ideas in algebraic
number theory. It is therefore of interest to repeat our analysis for a manifold of more than one
parameter.
\subsection{The mirror map and large complex structure limit}
Closely related to the issue of how to introduce \K\ parameters is the question of the mirror map and
the precise relation between the structure parameters of a manifold and the \K\ parameters of the
mirror. A proper understanding of the large (complex) structure limit is clearly also desirable.
\subsection{Modular properties of the $\z$-function}
We have seen that the $\z$-function for the conifold is related to a modular form and relations of this
type are known for rigid manifolds for which the motives are two dimensional. The question arises however
as to whether the $R$-factors that appear in the $\z$-function are also related to modular functions.
\subsection{The Euler curves}
This is an item of lesser significance since it relates most directly to the particular
family of manifolds that we have studied here. However a better understanding of the significance of the
Euler curves is also desirable. 
\vskip30pt
\leftline{\bf Acknowledgements}
It is a pleasure to acknowledge fruitful conversations with Benedict Gross, Roger Heath-Brown, Bong Lian,
Michael McQuillan,  John Tate, Felipe Voloch and Shin-Tung Yau. The work of FR-V is supported by a grant
of the NSF and TARP.
\newpage
\section{A}{Appendix A: The Large Complex Structure Limit}
We may regard $\ps$ as ranging over $\bb{P}\Fp$ so $\ps=\infty$ is an allowed value. There seems
however to be at least two ways to take this limit in the defining polynomial. One procedure is to
regard the defining polynomial as
 $$
P~=~u\left(x_1^5 + x_2^5 + x_3^5 + x_4^5 + x_5^5\right) - 5v\,x_1x_2x_3x_4x_5~,
~~~\hbox{with}~~~\ps={v\over u}~.$$
If we do this then the polynomial corresponding to $\ps=\infty$ is $\prod_i x_i=0$. This is the set of 
$x_i$ that have at least one coordinate zero. Thus 
 $$
N_r(\infty)~=~{(q^5-1)-(q-1)^5\over q-1}~=~5\,(p^r - p^{2r} + p^{3r})$$
which does not lead to an $\z$-function of the right form. Perhaps this should not surprise us since we
probably want to count the points of a resolution of the singular variety. One
resolution of this way of writing the singular manifold is its `normalization' which in this case gives five
disjoint hyperplanes. This would seem to give an $\z$-function of the form
 $$
\z_\ca{M}(t,\infty)~=~{1\over (1-t)^5(1-pt)^5(1-p^2t)^5(1-p^3t)^5}$$ 

In any event the process of resolving the singularity does not seem to be unique since we could also
proceed in the following way. Take the defining polynomial and make the change of coordinate, for
$\ps\in\Fpstar$, $5\ps x_1 = y$ then the polynomial becomes
 $$
P~=~\l\, y^5 + x_2^5 + x_3^5 + x_4^5 + x_5^5 - y\, x_2x_3x_4x_5 \eqlabel{singP}$$
and we can now take $\ps=\infty$ to correspond to $\l=0$. The variety corresponding to $\l=0$ is
singular at the single point $(y,0,0,0,0)$. One can construct the resolution of this singularity. It
will certainly be connected and so cannot consist of five disjoint
$\bb{P}_3$'s as previously. In any event consider now $\n_s(\infty)$ for the singular variety \eqref{singP}
over $\Fq$. Assume first that 
$5\notdiv q-1$ then we count by considering how many solutions there are with precisely $r$ of the four
coordinates $(x_2,x_3,x_4,x_5)$ equal to zero. 

If $r=0$ then we can rewrite the equation
 $$
y~=~{x_2^5 + x_3^5 + x_4^5 + x_5^5\over x_2x_3x_4x_5}$$ 
so there are $(q-1)^4$ solutions of this type. If $r=1$ then we have to count solutions of equations of the
form
 $$
x_i^5 + x_j^5 + x_k^5~=~0~,~~~i,j,k~\hbox{distinct}$$
with none of these coordinates zero. There are $4q\,\big((q-1)^2- (q-1)\big)$ of these. For $r=2$ there are
$6q(q-1)$ solutions to the equations $x_i^5 + x_j^5 = 0$. There are no solutions for $r=3$ and for $r=4$ we
have the $q$ solutions $(y,0,0,0,0)$. If we subtract 1 corresponding to the origin and divide out by
$(q-1)$ to account for the scaling of the coordinates we find
 $$
N_s(\infty)~=~q + q^2 + q^3~.$$
this leads to an $\z$-function of the form $1/\big( (1-pt)(1-p^2t)(1-p^3t)\big)$ which is not what we want
either.
\newpage
\section{B}{Appendix B: The Frobenius Period}
The purpose of this section is to review the general expression that Hosono, Lian and Yau
\cite{\HLY} give for the
Frobenius period in terms of the toric data. The point of immediate interest is that it makes clear the
fact that the $\e$ that has appeared in our use of the Frobenius method to construct solutions of the
Picard Fuchs equation for a manifold $\ca{M}$ really is the hyperplane class of $\bb{P}_4[5]$. More
generally the Frobenius period for a manifold $\ca{M}$ takes values in the even cohomology of the mirror.
Note that the total dimension of $H_{\bf  1}(\ca{W})\oplus H_2(\ca{W})\oplus H_4(\ca{W})\oplus H_6(\ca{W})$ is
$2h^{21}(\ca{M})+2$ so this gives the correct count to give a complete basis of periods for $\ca{M}$. 

Let $\ca{M}$ be a \cym\ given as a hypersurface in a toric variety $\bb{P}_\nabla$ by the zero locus of a
polynomial
 $$
P(x)~=~\sum_{\bm\in\D\atop\bm\neq\bone} a_\bm\, x^\bm - a_\bone x^\bone $$
with the convention that the coefficient, $a_\bone$, of the fundamental monomial is included with a
negative sign. It is shown in \cite{\HLY}
that the following period satisfies the GKZ system
 $$
\vp(a,D)~=~
{\prod_{\bm\in\D\atop\bm\neq\bone} \G(D_\bm + 1) \over \G(-D_\bone + 1)}
\sum_{\g\in V_\nabla} 
{\G(-\g\cdot D_\bone - D_\bone + 1) \over \prod_{\bm\in\D\atop\bm\neq\bone} \G(\g\cdot D_\bm + D_\bm + 1)}
\, a^{\g + D}~. \eqlabel{Fper}$$
In this expression the $D_\bm$ for $\bm\neq\bone$ are the toric divisors of $\bb{P}_\nabla$ that correspond
to the monomials $x^\bm$ while 
 $$
D_\bone~\define~ - [\ca{W}] = -\sum_{\bm\in\D\atop\bm\neq\bone} D_\bm$$ 
denotes minus the cohomology class of $\ca{W}$. The sum in \eqref{Fper} is over curves $\g$ in the Mori cone
of
$\bb{P}_\nabla$ with
$\g\cdot D_\bm$ denoting the intersection number of $\g$ with $D_\bm$. It is useful to think of $\g$ also
as a vector in the space dual to that spanned by the divisors. We extend this vector by prepending a
component $\g_\bone$ defined to be minus the sum of the remaining components. With this convention $\g$ has
length equal to the number of monomials $\bm$. We denote also by $D=(D_\bone,\ldots)$ a vector with
components $D_\bm$ and with these conventions
$a^{\g+D}$ denotes $\prod_{\bm\in\D} a_\bm^{\g_\bm + D_\bm}$. 
\subsection{The mirror quintic}
We want to revisit the periods for the mirror quintic. There are four of these and one complex structure
parameter. So for the present discussion $\ca{M}$ is the mirror quintic and $\D$ is the small polyhedron
with the five vertices and the interior point as its only points. We prepend a 1 to each $\bm$ and stack
the vectors to form a matrix as we have done before
 $$
\matrix{D_{\bf  1}\cr D_1\cr D_2\cr D_3\cr D_4\cr D_5\cr}
\pmatrix{
1&1&1&1&1&1\cr
1&5&0&0&0&0\cr
1&0&5&0&0&0\cr
1&0&0&5&0&0\cr
1&0&0&0&5&0\cr
1&0&0&0&0&5\cr}\eqlabel{mirrormatrix}$$
It is now simpler to label the divisors
$D_{\bf  1},D_1,\ldots,D_5$ with $D_{\bf  1}$ corresponding to the fundamental monomial. Written this way the divisors
$D_i$ with $i=0,\ldots,5$ correspond to the coordinate planes $x_i=0$ of $\ca{M}$. There are linear
relations between the divisors that we can read off from the columns of the matrix. The first column
provides the relation
$D_{\bf  1}=-\sum_{i=1}^5 D_i$ while the other columns tell us that the
$D_i,i=1,\ldots,5$ are all equal. Thus $D_i=H, i=1,\ldots,5$ and $D_{\bf  1}=-5H$. Since the \K\ cone is one
dimensional so is the Mori cone. The Mori cone lies in the null space of the matrix \eqref{mirrormatrix}
and it can be shown that $h=(-5,1,1,1,1,1)$ has the right sign. Thus $\g=nh, n=0,1,\ldots$. Return now to
the defining polynomial
 $$
P~=~a_1\,x_1^5 + a_2\,x_2^5 + a_3\,x_3^5 + a_4\,x_4^5 + a_5\,x_5^5 - a_{\bf  1}\,x_1x_2x_3x_4x_5$$
and make the change of variable $x_i=y_i/a_i^{1\over 5}$. This brings $P$ to the form
 $$
P~=~y_1^5 + y_2^5 + y_3^5 + y_4^5 + y_5^5 - 5\psi\, y_1y_2y_3y_4y_5$$
with $5\ps=a_{\bf  1}/(a_1a_2a_3a_4a_5)^{1\over 5}$ but then
 $$
\l~=~ {a_1a_2a_3a_4a_5\over a_{\bf  1}^5} ~=~ a^h~.$$
We have seen that $D=(-5,1,1,1,1,1)H$ though the fact that the same components appear here as in the curve
$h$ is a coincidence particular to this simple example. Finally we note that $\g\cdot D_i= n\,h\cdot H= n$
for $i=1,\ldots,5$ and $\g\cdot D_{\bf  1}= -5n$. Putting the pieces together we find
 $$
\vp(a,D)~=~
{\G(H + 1)^5 \over \G(5H + 1)}\sum_{n=0}^\infty {\G(5H + 5n + 1) \over \G(H + n + 1)^5}\, \l^{n+H}$$
which we indeed recognise as the Frobenius period.
\subsection{The mirror of the octic}
As a second example of the application of \eqref{Fper} we take the mirror of the manifold
$\bb{P}_4^{(1,1,2,2,2)}[8]$ that has two parameters and was studied in detail in~
\Ref{\CDFKM}{P.~Candelas, X.~del la Ossa, A.~Font and S.~Katz hep-th/9308083.}.
The mirror corresponds to a quotient $\bb{P}_4^{(1,1,2,2,2)}[8]/\G$ such that the most general octic is of
the form
 $$
P~=~y_1^8 + y_2^8 + y_3^4 + y_4^4 + y_5^4 - 2\ph\, y_1^4 y_2^4 - 8\ps\, y_1y_2y_3y_4y_5~.\eqlabel{octicP}$$
Now it was show in \cite{\CDFKM} that for $\bb{P}_4^{(1,1,2,2,2)}[8]$ there are two homology classes $H$
and $L$ that generate the \K\ cone and that
 $$
D_3 = D_4 = D_5 = H~~~\hbox{and}~~~D_1 = D_2 = L\eqlabel{Drels}$$
where the $D_i$  correspond to the vanishing of the coordinates $y_i$. We proceed as before with a matrix
that is now 
 $$
\matrix{D_{\bf  1}\cr D_1\cr D_2\cr D_3\cr D_4\cr D_5\cr D_6\cr}
\pmatrix{
1&1&1&1&1&1\cr
1&8&0&0&0&0\cr
1&0&8&0&0&0\cr
1&0&0&4&0&0\cr
1&0&0&0&4&0\cr
1&0&0&0&0&4\cr
1&4&4&0&0&0\cr}\eqlabel{mirrormatrix}$$
The columns of the matrix confirm \eqref{Drels} and tell us additionally that
 $$
D_{\bf  1}~=~-4H~~~\hbox{and}~~~D_6~=~2L - H~.$$
The null space of the matrix is two dimensional and we need to choose in this space a basis for the Mori
cone. It is straightforward to do this in terms of combinatorial properties of the matrix however since
this is not our main interest we can guess the answer by examining the large complex structure coordinates
which for this case are
 $$
\l~=~-{2\ph\over (8\ps)^4}~~~\hbox{and}~~~\m~=~{1\over (2\ph)^2}~.$$
By writing out the general form of the mirror octic
 $$
P~=~a_1\, x_1^8 + a_2\, x_2^8 + a_3\, x_3^4 + a_4\, x_4^4 + a_5\, x_5^4 + a_6\, x_1^4 x_2^4 - 
a_{\bf  1}\, x_1x_2x_3x_4x_5$$ 
and scaling coordinates we find that
 $$
\l~=~{a_3a_4a_5a_6\over a_{\bf  1}^4}~~~\hbox{and}~~~\m~=~{a_1a_2\over a_6^2}$$
So we guess (correctly) that the generators are given by the exponents in these relations
$(\l,\m)=(a^h,a^l)$ with
 $$
h~=~(-4,0,0,1,1,1,1)~~~\hbox{and}~~~l~=~(0,1,1,0,0,0,-2)~.$$
Setting now $\g=nh+ml$ we find \eqref{Fper} now reduces to 
 $$\eqalign{
\vp(a,D)~&=~{\G(H+1)^3\, \G(L+1)^2\, \G(H-2L+1) \over \G(4H+1)}\cropen{10pt}
&\hskip10pt\times\sum_{n,m=0}^\infty {\G(4H+4n+1) \over \G(H+n+1)^3\, \G(L+m+1)^2\, \G(H-2L+n-2m+1)} 
\,\l^{n+H}\m^{m+L}~.\cr}$$
To make contact with the fundamental period given in \cite{\CDFKM} we set $H=L=0$ in this expression and
note that, owing to the factor of $\G(n-2m+1)$ in the denominator, the resulting summand vanishes for
$n<2m$. Hence
 $$\eqalign{
\vp_0(\l,\m)~&=~\sum_{m=0}^\infty\sum_{n=2m}^\infty {(4n)!\over (n!)^3 (m!)^2 (n-2m)!}
\,\l^n\m^m\cropen{5pt}  
 &=~\sum_{m=0}^\infty \sum_{s=0}^\infty {(8m+4s)!\over \big((2m+s)!\big)^3
(m!)^2 s!}\, {(-2\ph)^s \over (8\ps)^{8m+4s}}\cr}$$
Where the last line follows on making the change of variable $n=2m+s$ and is precisely the period given in
\cite{\CDFKM}.
\newpage
\frenchspacing
\immediate\closeout\referencewrite\referenceopenfalse
\line{\fourteenbold\hfil References\hfil}\bigskip\parindent=0pt\input referenc.texauxil
\bye